\documentstyle[12pt]{article}
\newfont{\bg}{cmr10 scaled\magstep4}

\newcommand{\bigzerou}{%
   \smash{\lower1.7ex\hbox{\bg 0}}}
\def\a{\begin{eqnarray}}
\def\b{\end{eqnarray}}
\def\0{\nonumber}
\def\ba{\begin{array}}
\def\ea{\end{array}}
\begin{document}
\title{
\begin{flushright}
  \begin{minipage}[b]{5em}
    \normalsize
      UT-HEP-703-95\\
  \end{minipage}
\end{flushright}
Construction of Free Energy of Calbi-Yau manifold\\
 embedded in $CP^{N-1}$ via Torus Actions}
\author{Masao Jinzenji \\
\\
\it Department of Phisics, University of Tokyo\\
\it  Bunkyo-ku, Tokyo 113, Japan}
\maketitle
\begin{abstract}
We calculate correlation functions of topological sigma model (A-model)
on C.Y.hypersurfaces in $CP^{N-1}$ using torus action method.
We also obtain
path-integral representation of free energy of the theory coupled to
gravity.
\end{abstract}

\section{Introduction}
Recently vast development occurs in the field of topological sigma
model, especially from the mirror symmetry of the models having
Calabi-Yau manifolds as target spaces \cite{cogp}.
  In \cite{nj}, Nagura and the author treated the topological sigma
model
on the Calabi-Yau
hypersurfaces $M_{N}$ in $CP^{N-1}$. We derived $N-2$ point
correlation
function by the analysis of the solution of period integral equation
which emerges from the deformation of the complex structure of
$\tilde{M}_{N}$,the mirror counterpart of $M_{N}$(B-model).
And we showed that translation of the calculation of B-model
into the A-model by mirror map actually
gives the correlation function of topological sigma model on
$M_{N}$, i.e.,
the number of holomorphic maps satisfying the conditions imposed by
external operator insertion. But the explicit evaluation was limited
only
for the degree $1$ case. So there remains a problem of the calculation of
the correlation functions from the A-model point of view for higher degree
cases. Of course, search for the deeper understanding of mirror symmetry is
very important, but we limit our interest to the above problem in this
paper.

As we saw in \cite{nj}, correlation functions of the A-model have geometrically
very clear meaning. And their direct calculation relies on the explicit
construction of moduli spaces for holomorphic maps from $CP^{1}$ to target
space (in our case $M_{N}$). For degree 1, this can be done by taking
the zero locus of a section of $Sym^{N}(U^{*})$ of $Gr(2,N)$ ($U$ is the
 universal bundle of Grassmannian). S.Katz constructed degree $2$ moduli
space as the zero locus of a section of $Sym^{N}(U^{*})/Sym^{N-2}(U^{*})
\otimes {\cal O}_{P}(-1)$ on $Gr(3,N)$.(${\cal O}_{P}(1)$ is the
taughtological
sheaf on $P(Sym^{2}(U))$.) These construction seems to become more complicated
as the degree rises. But in \cite{tor}, Kontsevich constructed the compact
moduli
space from $CP^{1}$ to complex manifold $V$ by introducing stable maps.
Roughly speaking, these are maps from branched $CP^{1}$ (stable
curves) to $V$. In the case where $V$ is $CP^{N}$ or hypersurfaces of
$CP^{N}$, he also did some calculations of correlation functions
(Gromov-Witten invariants) using this construction and by means of fixed
point theorem under the torus action flow of $CP^{N}$. We find that his
treatment of quintic hypersurface in $CP^{4}$ is very much like our
elementary approach for $M_{N}$. So we thought we can calculate
Gromov-Witten
invariants for $M_{N}$ by this method.

We have to notice one difference. In \cite{nj},we treated the matter theory,
but
in Kontsevich's formulation correlation functions are those for the theory
coupled with gravity because of introduction of stable maps. After all, we
found this method also works for $M_{N}$. Using the fact that $3$-point
functions are identical for both theories and that fusion rule holds in
the matter theory (See Greene,Morrison,Plesser \cite{gmp}), we reconstruct
$N-2$ point
functions for matter theory and derived some identity.

This method has a by-product. It has the structure of the sum of tree
graph amplitudes, so by using Feynmann rules, we can construct path-integral
representation of the generating function of correlation functions (Free
Energy). We will represent it at the end of this paper.

In section $2$, we introduce the topological sigma model (A-model) and
construct correlation functions as integrals of forms on moduli spaces.
In section $3$, we review the torus action method and in section $4$, we
do some explicit calculation of amplitudes and see these results are
compatible with those of Greene, Morrison, Plesser \cite{gmp},and
S.Katz \cite{katz}. In section 5
we construct path-integral representation of free energy for $M_{N}$
(coupled with gravity).

This paper's treatment is rather non-rigolous in comparison with the
original one of Kontsevich, but we think it is more accessible and may arise
some insight for generalization.
\section{Correlation Function as an Integral of Forms on Moduli Spaces}
\subsection{Topological Sigma Model (A-Model)}
\subsubsection{lagrangian and weak coupling limit}

 Topological sigma model can be
obtained by twisting $N=2$ Supersymmetric Sigma Model on M. A-model
corresponds to A-twist, which turns $\psi_{+}^i$ and $\psi_{-}^{\bar{i}}$
in $N=2$ Sigma Model into $\chi^{i}$,$\chi^{\bar{i}}$, and
$\psi_{+}^{\bar{i}},\psi_{-}^i$
into $\psi_{z}^{\bar{i}},\psi_{\bar{z}}^{i}$. In other words, A-twist means
subtraction of half of $U(1)$ charge from world sheet spin quantum number.
Then we obtain the following Lagrangian for the A-model,

\begin{equation}
 L = 2t \int_{\Sigma} d^{2}z(\frac{1}{2}g_{IJ}\partial_{z}\phi^{I}
\partial_{\bar{z}}\phi^{J} + i\psi_{z}^{\bar{i}}D_{\bar{z}}\chi^{i}
g_{i\bar{i}} + i\psi_{\bar{z}}^{i}D_{z}\chi^{\bar{i}}g_{\bar{i}i}
-R_{i\bar{i}j\bar{j}}\psi_{\bar{z}}^{i}\psi_{z}^{\bar{i}}\chi^{j}
\chi^{\bar{j}})
\label{a1}
\end{equation}

(\ref{a1}) is invariant under the
BRST - transformation,

\begin{eqnarray}
   & \delta\phi^{i} = i\alpha\chi^{i} \nonumber\\
   & \delta\phi^{\bar{i}} = i\alpha\chi^{\bar{i}} \nonumber\\
   & \delta \chi^{i} = \delta \chi^{\bar{i}} = 0 \nonumber\\
   & \delta\psi_{z}^{\bar{i}} = -\alpha\partial_{z}\phi^{\bar{i}}
         -i\alpha\chi^{\bar{j}}\Gamma^{\bar{i}}_{\bar{j}\bar{m}}
         \psi_{z}^{\bar{m}}
        \nonumber\\
   & \delta\psi_{\bar{z}}^{i} = -\alpha\partial_{\bar{z}}\phi^{i}
         -i\alpha\chi^{j}\Gamma^{i}_{jm}\psi_{\bar{z}}^{m}
\label{a2}
\end{eqnarray}

This invariance allows us to consider only BRST-invariant observables.
As in section2, we define BRST operator $Q$ by $\delta V =
 -i\alpha\{Q,V\}$ for any field V. Of course, $Q^2 = 0$.
 We can rewrite the lagrangian (\ref{a1}) modulo the $\psi$ equation of
motion as;
\begin{equation}
  L = it\int_{\Sigma}d^2z \{Q,V\} + t\int_{\Sigma}\Phi^{*}(e)
\label{a3}
\end{equation}
where
\begin{equation}
V = g_{i\bar{j}}(\psi_{z}^{\bar{i}}\partial_{\bar{z}}\phi^{j} + \partial_{z}
     \phi^{\bar{i}}\psi_{\bar{z}}^{j})
\label{a4}
\end{equation}
and
\begin{equation}
 \int_{\Sigma}\Phi^*(e) = \int_{\Sigma}(\partial_{z}\phi^{i}\partial_{\bar{z}}
     \phi^{\bar{j}}g_{i\bar{j}}- \partial_{\bar{z}}\phi^{i}
     \partial_{z}\phi^{\bar{j}}g_{i\bar{j}})
\label{a5}
\end{equation}
(\ref{a5}) is the integral of the pullback of the K\"ahler form $e$ of M,
and it
depends only on the intersection number between $\Phi_{*}(\Sigma)$ and
$PD(e)$($PD(e)$ denotes the Poincare Dual of $e$), which
equals the degree of $\Phi$. By an appropriate normalization
 of $g_{i\bar{j}}$,we have
\begin{equation}
\int_{\Sigma}\Phi^{*}(e) = n
\label{a60}
\end{equation}
where n is the degree of $\Phi$.
 Next, we consider the correlation function of BRST-invariant observables
$\{ {\cal O}_{i}\}$, i.e.
\begin{equation}
\langle \prod_{i=1}^{k} {\cal O}_{i}\rangle = \int {\cal D}\phi
 {\cal D}\psi{\cal D}\chi e^{-L} \prod_{i=1}^{k} {\cal O}_{i}
\label{a6}
\end{equation}
 We have seen $\int_{\Sigma}\Phi^{*}(e) = n $ and we decompose the
space of maps $\phi$ into different topological sectors
$\{B_{n}\}$ in each of which $deg(\Phi)$ is a fixed
integer.

 We can rewrite (\ref{a6}) as follows.
\begin{equation}
\langle \prod_{i=1}^{k} {\cal O}_{i}\rangle = \int {\cal D}\phi
 {\cal D}\psi{\cal D}\chi e^{-L} \prod_{i=1}^{k} {\cal O}_{i}
     = \sum_{n=0}^{\infty}e^{-nt}\int_{B_n} {\cal D}\phi
 {\cal D}\psi{\cal D}\chi e^{-it\int_{\Sigma}d^2 z \
    \{Q,V\}} \prod_{i=1}^{k} {\cal O}_{i}
\label{a7}
\end{equation}
 And we set
\begin{equation}
\langle \prod_{i=1}^{k} {\cal O}_{i}\rangle_{n}
   = \int_{B_n} {\cal D}\phi
 {\cal D}\psi{\cal D}\chi e^{-it\int_{\Sigma}d^2 z \
    \{Q,V\}} \prod_{i=1}^{k} {\cal O}_{i}
\label{a8}
\end{equation}
 We can easily see that $\int_{\Sigma}d^{2}z\{Q,V\} = \{Q,\int_{\Sigma}d^{2}
zV\}$,i.e. lagrangian is BRST exact. It follows from this and $\{Q,{\cal O}_i
\}=0$ that $\langle \prod_{i=1}^{k}{\cal O}_i\rangle_{n}$ doesn't depend on the
coupling constant t and we can take weak coupling limit
$t \rightarrow \infty $ in evaluating the path integral.

 In this limit,the saddle point approximation of the path integral
 becomes exact. Saddle points of the lagrangian are given by
\begin{equation}
 \partial_{\bar{z}}\phi^{i} = 0\;\;\; \partial_{z}\phi^{\bar{i}} = 0
\label{a9}
\end{equation}
 (\ref{a9}) shows that the path-integral is reduced to an integral over
 the moduli space
 of holomorphic maps from $\Sigma$ to $M$ of degree $n$.  We will focus
our attention to the
case where $\Sigma$ has  genus 0.  We denote
this space as ${\cal M}_{0,n}^{M}$.
\subsubsection{The Ghost Number anomally and BRST observables }
 In the previous subsection, we have seen that the path integral
(\ref{a6}) is reduced to an integral over ${\cal M}_{0,n}^{M}$
weighted by one loop
determinants of the non zero modes. But in general, there are fermion zero
modes  which are given as the solution of $D_{\bar{z}}\chi^{i}=
D_{z}\chi^{\bar{i}}=0$ and $D_{\bar{z}}\psi_{z}^{\bar{i}}=
D_{z}\psi_{\bar{z}}^{i} = 0$. Let $a_n$ (resp.$b_n$) be the number of $\chi$
(resp.$\psi$) zero modes. If $M$ is a Calabi-Yau manifold, we can see from
Riemann-Roch Theorem,
\begin{equation}
 w_n=a_n-b_n=\mbox{dim}(M)
\label{a10}
\end{equation}
 The existence of fermion zero mode is understood as Ghost number anomally,
because lagrangian (\ref{a1}) classically conserves the ghost number.
  In path integration,these zero modes appear only in the integration measure
except in $\prod_{i=1}^k{\cal O}_i$,and the correlation function
$\langle \prod_{i=1}^k{\cal O}_i
\rangle_{n}$ vanishes unless the sum of
the ghost number of ${\cal O}_i$ is equal
to $w_n$.

 $D_{\bar{z}}\chi^{i}=0$ (resp.$D_{z}\chi^{\bar{i}}=0$) can be considered as
the linearization of the equation $\partial_{\bar{z}}\phi^{i}=0$
(resp.$\partial_{z}
\phi^{\bar{i}}=0$) and we can regard $\chi$ zero mode as
$T{\cal M}_{0,n}^{M}$.
$w_n$ is usually called ``virtual dimension `` of ${\cal M}_{0,n}^{M}$.
 In generic case $b_n=0$ and $\mbox{dim}({\cal M}_{0,n}^{M}) = w_n$
holds.
Then we have $\mbox{dim}({\cal M}_{0,n}^{M})=a_n$.
 BRST cohomology classes of the A-model are constructed from the de Rham
cohomology classes $H^{*}(M)$ of the manifold M.
 Let $W= W_{I_1I_2\cdots I_n}(\phi)d\phi^{I_1}\wedge\cdots\wedge d\phi^{I_n}$
 be an $n$ form on $M$. Then we define a corresponding local operator of the
A-Model,
\begin{equation}
{\cal O}_{W}(P) = W_{I_1\cdots I_n}\chi^{I_1}\cdots\chi^{I_n}(P)
\label{a11}
\end{equation}
 From (\ref{a2}) we have
\begin{equation}
\{Q,{\cal O}_{W}\}=-{\cal O}_{dW}
\label{a12}
\end{equation}
which shows that if $W \in H^{*}(M)$,${\cal O}_{W}(P)$ is BRST-closed.
\subsubsection{Evaluation of the Path Integral}
 Now we discuss how we can evaluate $\langle \prod_{i=1}^{k}{\cal O}_i \rangle
_n$ . We take ${\cal O}_i$ to be
${\cal O}_{W_i}$ which is induced from $W_i \in H^{*}(M)$. By adding
appropriate exact forms  we can make $W_i$ into the differential form
which has delta function support on $PD(W_i)$. Then
${\cal O}_{W_i}(P_i)$ is non zero only if
\begin{equation}
 \phi(P_i) \in PD(W_i)
\label{a13}
\end{equation}
Then integration over ${\cal M}_{0,n}^{M}$ is restricted to
$\tilde{\cal M}_{0,n}^{M}$
, which consists of $\phi \in {\cal M}_{0,n}^{M}$ satisfying
(\ref{a13}). In
evaluating $\langle \prod_{i=1}^{k}{\cal O}_{W_i}\rangle_n$, (\ref{a13})
imposes $\sum_{i=1}^{k}\mbox{dim}(W_i)$ conditions, so
$\mbox{dim}(\tilde{\cal M}_{0,n}^{M}) =
\mbox{dim}({\cal M}_{0,n}^{M})- \sum_{i=1}^{k}\mbox{dim}(W_i) =
w_n + b_n -\sum_{i=1}^{k}\mbox{dim}(W_i)$.
But from the fact that ghost number of ${\cal O}_{W_i}$ equals
$\mbox{dim}(W_i)$
(contribution from $\chi$) and anomally canselation condition,we have
$\mbox{dim}(\tilde{\cal M}_{0,n}^{M}) = b_n$.
In generic case where $b_n=0$,
$\tilde{\cal M}_{0,n}^{M}$
turns into finite set of points. Then we perform an one loop integral
over each of these points. The result is a ratio of boson and fermion
determinants,which cancel each other. Then contributions to
$\langle \prod_{i=1}^{k} {\cal O}_{W_i} \rangle_{n}$
in the generic case equals
the number of instantons which satisfies (\ref{a13}),i.e.
\begin{equation}
\langle \prod_{i=1}^{k} {\cal O}_{W_i} \rangle_{generic} =
{}^{\sharp} \tilde{\cal M}_{0,n}^{M}
\label{a14}
\end{equation}
 When $dim(\tilde{\cal M}_{0,n}^{M}) = b_n \geq 1$,
there are $b_n$ $\psi$ zero modes
which we can regard as the fiber of the vector bundle $\nu$ on $\tilde{\cal
M}_{0,n}^{M}$.
In this case, contributions to
$\langle \prod_{i=1}^{k} {\cal O}_{W_i} \rangle_{n}$are known as the
integration of Euler class $\chi(\nu)$ on $\tilde{\cal M}_{0,n}^{M}$.
If we consider
$\nu$ as a 0-dimensional vector bundle on a point in the generic case, we can
apply the same logic there. We denote each component of
$\tilde{\cal M}_{0,n}^{M}$ as
$\tilde{\cal M}_{0,n,m}^{M}$ and obtain
\begin{equation}
\langle \prod_{i=1}^{k} {\cal O}_{W_i} \rangle_{n} = \sum_{m}
\int_{\tilde{\cal M}_{0,n,m}^{M}}\chi(\nu)
\label{a15}
\end{equation}
Hence from (\ref{a7})
\begin{equation}
\langle \prod_{i=1}^{k} {\cal O}_{W_i} \rangle = \sum_{n=0}^{\infty}
\sum_{m=1}^{m_n}e^{-nt}\int_{\tilde{\cal M}_{0,n,m}^{M}}\chi(\nu)
\label{a16}
\end{equation}
 In algebraic geometry, generic instantons of degree $n$ corresponds to
irreducible maps of degree n, and instantons of degree n with non-zero $\psi$
 zero mode to reducible maps which are $j$-th multiple cover of irreducible
maps of degree $n/j$ $(j|n)$.
 Let $\tilde{\cal M}_{0,n,j,m}^{M}$ be the $m$-th connected component
of moduli
spaces which are $j$-th multiple cover of $n/j$-th irreducible instantons,
 and $\nu_{j,m}$ be vector bundle of $\psi$ zero modes on
$\tilde{\cal M}_{0,n,j,m}^{M}$. Then we have from (\ref{a16}),
\begin{equation}
\langle \prod_{i=1}^{k} {\cal O}_{W_i} \rangle = \sum_{n=0}^{\infty}
\sum_{j|n}\sum_{m=1}^{m_{n,j}}e^{-nt}\int_{\tilde{\cal M}_{0,n,j,m}^{M}}
\chi(\nu_{j,m})
\label{a17}
\end{equation}
\subsection{Reduction to an integral of Forms on Moduli Spaces}
In this paper, we treat the topological sigma model (A-Model) of Calabi-Yau
manifold embedded in $CP^{N-1}$. This manifold is realized as  the zero-locus
of
section of $-K$ ($K$ is canonical line bundle of $CP^{N-1}$.).Since $-K$ is
equivalent to $ N H $ as a line bundle ($H$ is hyperplane bundle of
$CP^{N-1}$), we can take homogeneous polynomial of degree $N$ as
the defining
equation of $M_{N}$. For example,
\begin{equation}
 M_{N} := \{(X_{1},X_{2},\cdots ,X_{N}) / \mbox{\bf C}^{\times} \in CP^{N-1}|
 X_{1}^{N}+X_{2}^{N}+\cdots + X_{N}^{N}=0 \}
\end{equation}
Observables of this model can be constructed from elements of
$w \in H^{*}(M_{N})$
which we denote as ${\cal O}_{w}$, and in the following discussion
we consider the
observables which are included from the subring of $H^{*}(M_{N},C)$ generated
by
K\"ahler form $e$ of $M_{N}$ (we denote it as $H_{CP^{N-1}}(M_{N},C)$). One of
the
reason why we take this subring is that we can obtain it directly from
$H^{*}(CP^{N-1},C)$ and Poincare dual of its elements are
analytic submanifold of $M_{N}$. More explicitly, elements of
$H_{CP^{N-1}}(M_{N},C)$
are given as $e^{k}$ ($k = 1,2,\cdots ,N-2 $) and Poincare dual of $e^{k}$
is the intersection of
the zero locus of the section of  $H^{0}(CP^{N-1},{\cal O}(k\cdot H))$ and
$M_{N}$.
 So in the following discussion we treat the observables
\footnote{When coupled to
gravity,${\cal O}_{1}$ correspond to puncture operator $P$ but in the small
phase,
space $P$ insertion is suppressed except for constant map sector because of
puncture equation and as we know from the later discussion of topological
selection
rule, ghost number of inserted operator must be less than $N-3$. So it
suffices to
consider only $N-4$ elements
${\cal O}_{e},{\cal O}_{e^2},\cdots,{\cal O}_{e^{N-4}}$.}
\begin{equation}
 {\cal O}_{1},{\cal O}_{e},{\cal O}_{e^{2}},\cdots ,{\cal O}_{e^{N-2}}
\end{equation}
Then the fact that Lagrangian of the topological sigma model is BRST-exact
allows us
to take the strong coupling limit and correlation functions of this model
reduces to
the integral of closed forms corresponding to the BRST closed observables on
moduli
spaces of holomorphic maps $f$ from Riemann surface $\Sigma_{g}$ to target
space
$M_{N}$ (we focus our attention to the case of $g=0$,i.e,$CP^{1}$).
When the
target space
is a hypersurface of simple projective space $CP^{N-1}$,we can classify
moduli spaces
by the degree $n= {}^{\sharp}(f(CP^{1})\cap PD(e))$ and we denote the
moduli space
of degree $n$ as ${\cal M}^{M_{N}}_{0,n}$. Dimension of
${\cal M}^{M_{N}}_{0,n}$
which counts the number of $\chi$-zero modes is evaluated by the
Riemann-Roch
Theorem as follows.
\begin{eqnarray}
\mbox{dim}({\cal M}^{M_{N}}_{0,n})&:=&
\mbox{dim}(H^{0}(CP^{1},f^{*}(T^{\prime}M_{N})))
\nonumber\\
                 &=& \mbox{dim}(M_{N}) + {\mbox deg}(f) \cdot c_{1}(KM_{N})
                         + \mbox{dim}(H^{1}(CP^{1},f^{*}(T^{\prime}M_{N})))
\nonumber\\
                 &=& \mbox{dim}(M_{N}) +
\mbox{dim}(H^{1}(CP^{1},f^{*}(T^{\prime}M_{N})))
\nonumber\\
                 &=& N-2 + \mbox{dim}(H^{1}(CP^{1},f^{*}(T^{\prime}M_{N})))
\label{roch}
\end{eqnarray}
 where we used the Calabi-Yau condition $c_{1}(K_{M_{N}})=0$.(\ref{roch})
tells
us that the
dimension of moduli space is independent of degree $n$.

 First, we consider the generic case where
$\mbox{dim}(H^{1}(CP^{1},f^{*}(T^{\prime}M_{N})))=0$.
{}From the above argument, we can heuristically represent correlation
functions,
\begin{eqnarray}
\langle {\cal O}_{e^{j_{1}}}(z_{1}){\cal O}_{e^{j_{2}}}(z_{2}) \cdots
{\cal O}_{e^{j_{k}}}(z_{k}) \rangle _{n,generic}\nonumber\\
 = \int_{{\cal M}^{M_{N}}_{0,n}} \alpha({\cal O}_{e^{j^{1}}})\wedge
\alpha({\cal O}_{e^{j^{2}}})
\wedge \cdots \wedge  \alpha({\cal O}_{e^{j^{k}}})
\end{eqnarray}
where $ \alpha({\cal O}_{e^{j}})$ is the closed form on
${\cal M}^{M_{N}}_{0,n}$ induced
from ${\cal O}_{e^{j}}$. Since the form degree of  $ \alpha({\cal O}_{e^{j}})$
equals the ghost
number of ${\cal O}_{e^{j}}(=\mbox{dim}(e^{j})=j)$,correlation functions
are nonzero
only if the
following conditions are satisfied.
\begin{eqnarray}
 \mbox{dim}({\cal M}^{M_{N}}_{0,n})&=& \sum_{i=1}^{k} j_{i} \nonumber\\
\Longleftrightarrow N-2 &=&  \sum_{i=1}^{k} j_{i}
\label{sel}
\end{eqnarray}
If we take $e^{j}$ as the forms which has the delta function support on
$PD(e^{j})$,then from ( ),
$\alpha ({\cal O}_{e^{j}})$ can be interpreted as the constraint condition
on $f$,
\begin{equation}
 \alpha ({\cal O}_{e^{j_{i}}}) \leftrightarrow f(z_{i}) \in PD(e^{j_{i}})
\label{con}
\end{equation}
(\ref{con}) imposes $({\mbox dim}(e^{j})-1)+1 = j$ independent conditions on
${\cal M}^{M_{N}}_{0,n}$
($(\mbox{dim}(e^{j})-1)$ corresponds to the degree of freedom which makes
$f(CP^{1}) \cap PD(e^{j})\neq \emptyset$ and $1$ to the one which sends
$f(z_{i})$ into $PD(e^{j})$). And from (\ref{sel}),
what remains is the discrete point set of holomorphic maps $f$ which
satisfy (\ref{con}) for
all $i$. Then we have
\begin{eqnarray}
\langle {\cal O}_{e^{j_{1}}}(z_{1}){\cal O}_{e^{j_{2}}}(z_{2}) \cdots
{\cal O}_{e^{j_{k}}}(z_{k}) \rangle _{n,generic}\nonumber\\
 = {}^{\sharp}\{f : CP^{1} \stackrel{hol.}{\rightarrow}
M_{N} |f(z_{i}) \in PD(e^{j_{i}})\}
\end{eqnarray}
 Now, let us consider non-generic case.In this case,
$\mbox{dim}(H^{1}(CP^{1},f^{*}(T^{\prime}M_{N}))>0$
and moduli space have additional
$\mbox{dim}(H^{1}(CP^{1},f^{*}(T^{\prime}M_{N}))$
degrees of freedom.

We can see these degrees of freedom correspond to multiple cover
maps by
the
following argument.
A multiple cover map $f$ can be decomposed into the form
$f = \tilde{f}\circ \varphi$ where $\tilde{f}$
is irreducible map from $CP^{1}$ to $M_{N}$ and $\varphi$ represents the map
from $CP^{1}$ to
$CP^{1}$ of degree $d \geq 2$. Then let us  count
$\mbox{dim}({\cal M}^{M_{N}}_{0,n})$ by taking the
(holomorphic) variation of $\tilde{f} \circ \varphi$.
\begin{eqnarray}
    \delta(\tilde{f}\circ\varphi)\nonumber\\
   = \delta\tilde{f}\circ\varphi + \tilde{f}\circ\delta\varphi
\end{eqnarray}
$\delta\tilde{f}$ corresponds to the generic degrees of freedom,i.e,
\begin{equation}
 \mbox{dim}(\delta\tilde{f}\circ\varphi)= N-2
\end{equation}
$\tilde{f}\circ\delta\varphi$ counts the deformation of the multiple cover
map which can be
realized using the section
$\varphi^{z} \in H^{0}(CP^{1},\varphi^{*}(T^{\prime}CP^{1}))$ as
$\varphi^{z}\partial_{z}\tilde{f}$. We can count these by using
Riemann-Roch,
\begin{eqnarray}
\mbox{dim}(\tilde{f}\circ\delta\varphi)&=&
\mbox{dim}(H^{0}(CP^{1},\varphi^{*}(T^{\prime}CP^{1})))-3\nonumber\\
                  &=&1 + {\mbox deg}(\varphi)\cdot c_{1}(T^{\prime}CP^{1})-3
\nonumber\\
                  &=& 2d -2
\label{multi}
\end{eqnarray}
In (\ref{multi}) we subtract the double counted $SL(2,C)$
which comes from the indetermination
of the decomposition of $f$,i.e,
\begin{eqnarray}
    f &=& \tilde{f}\circ\varphi \nonumber\\
      &=& \tilde{f}\circ u \circ u^{-1} \circ \varphi \qquad  u \in SL(2,C)
\end{eqnarray}
 After all we find additional $2d-2$ $\chi$ zero modes.But we can also
construct $2d-2$ $\psi_{\bar{z}}^{i}$
which comes from $H^{1}(CP^{1},f^{*}(T^{\prime}M_{N}))$. By the Kodaira-Serre
duality ,the following equation holds.
\begin{eqnarray}
\mbox{dim}(H^{1}(CP^{1},f^{*}(T^{\prime}M_{N})) &=&
\mbox{dim}(H^{0}(CP^{1},K\otimes f^{*}(T^{\prime *}M_{N})))\nonumber\\
 (&=& \mbox{dim}(H^{0}(CP^{1},\bar{K}\otimes f^{*}(\bar{T}^{\prime *}M_{N}))))
\end{eqnarray}
and
\begin{equation}
    g^{i \bar{j}}\tilde{\psi}_{\bar{z},\bar{j}}
\qquad (\tilde{\psi}_{\bar{z},\bar{j}} \in H^{0}(CP^{1},\bar{K}\otimes
f^{*}(\bar{T}^{*,\prime}M_{N})))
\end{equation}

 Lagrangian (\ref{a1}) has the
$R_{i \bar{i} j \bar{j}}\psi_{\bar{z}}^{i}
\psi_{z}^{\bar{i}}\chi^{j}\chi^{\bar{j}}$ term,
so we can ``kill'' additional $2d-2$ $\chi$ and $\psi$ zero modes by
expanding
$\exp(R_{i \bar{i} j \bar{j}}
\psi_{\bar{z}}^{i}\psi_{z}^{\bar{i}}\chi^{j}\chi^{\bar{j}})$.
So we don't have to
add extra operator insertions. Then,by integrating $\psi$ zero-modes first,
we have the Euler class
$\chi(\nu)$ where $ \nu \simeq H^{1}(CP^{1},\varphi^{*}(T^{\prime}M_{N}))$.
This leads to

\begin{eqnarray}
\langle {\cal O}_{e^{j_{1}}}(z_{1}){\cal O}_{e^{j_{2}}}(z_{2}) \cdots
{\cal O}_{e^{j_{k}}}(z_{k}) \rangle _{n} \nonumber\\
= \int_{{\cal M}_{0,n}^{M_{N}}} \chi(\nu)
\wedge \alpha({\cal O}_{e^{j_{1}}}(z_{1})) \wedge
\cdots \alpha({\cal O}_{e^{j_{k}}}(z_{k}))
\label{13}
\end{eqnarray}
 We can refine (\ref{13}) by using the argument which leads to (\ref{con})
and define the evaluation map,
\begin{equation}
 \varphi_{i} : {\cal M}_{0,n}^{M_{N}} \rightarrow M_{N}:f \mapsto f(z_{i})
\end{equation}
We have
\begin{eqnarray}
\langle {\cal O}_{e^{j_{1}}}(z_{1}){\cal O}_{e^{j_{2}}}(z_{2}) \cdots
{\cal O}_{e^{j_{k}}}(z_{k}) \rangle _{n} \nonumber\\
= \int_{{\cal M}_{0,n}^{M_{N}}} \chi(\nu)
\wedge \varphi^{*}_{1}(e^{j_1}) \wedge \cdots \varphi^{*}_{k}(e^{j_{k}})
\label{15}
\end{eqnarray}
We can relate the non-generic part the correlation functions to
 ones of
lower degree, because in
such case $f$ decomposes into $f= \tilde{f}\circ\varphi$ where
$\mbox{deg}(\varphi)=d$ and $\mbox{deg}(\tilde{f})=
n/d < N$. But good results are given only in the case of
$k=3$, which
was derived by Greene, Aspinwall, Morrison
 and Plesser \cite{gmp} \cite{am}. Of course, if we use the fusion rule that
holds in
the matter theory, we can
reduce the correlation functions into the product of three point
functions
and formally distinguish
the non-generic part from the generic ones.But geometrical meaning is still
not clear.

 Then we slightly change our point of view. Since $M_{N}$ is a hypersurface
in $CP^{N-1}$,we can see
${\cal M}_{0,n}^{M_{N}}$ as a submanifold of ${\cal M}_{0,n}^{CP^{N-1}}$
which consists of maps
satisfying the following condition.
\begin{eqnarray}
    f: CP^{1} \rightarrow CP^{N-1}\nonumber\\
    f(z) \in M_{N} \qquad \mbox{for all} \quad z \in CP^{1}
\label{16}
\end{eqnarray}
If we can realize the condition (\ref{16}) as the closed forms
$c_{n}(M_{N})$ on
${\cal M}_{0,n}^{CP^{N-1}}$, we have an alternate representation for the
correlation functions
as follows,
\begin{eqnarray}
\langle {\cal O}_{e^{j_{1}}}(z_{1}){\cal O}_{e^{j_{2}}}(z_{2}) \cdots
{\cal O}_{e^{j_{k}}}(z_{k}) \rangle _{n,alt} \nonumber\\
= \int_{{\cal M}_{0,n}^{CP^{N-1}}} c_{n}(M_{N})
\wedge \tilde{\varphi}^{*}_{1}(e^{j_1}) \wedge \cdots \wedge
\tilde{\varphi}^{*}_{k}(e^{j_{k}})\nonumber\\
 \tilde{\varphi}_{i} : {\cal M}_{0,n}^{M_{N}}
\rightarrow CP^{N-1}:f \mapsto f(z_{i})
\label{17}
\end{eqnarray}

In (\ref{17}), we can drop off the Euler class $\chi(\nu)$.
This is because
\begin{eqnarray}
\mbox{dim}(H^{1}(CP^{1},\tilde{\varphi}^{*}(T^{\prime}CP^{N-1})))
&=&\mbox{dim}(H^{0}(CP^{1},K\otimes(\tilde{\varphi}^{*}
(T^{\prime}M_{N}))))
=0
\nonumber \\
(c_{1}(K_{CP^{1}}\otimes \varphi^{*}(K_{CP^{N-1}}))
&=& \mbox{deg}(\varphi) \cdot (-2) \cdot N < 0)
\end{eqnarray}
 Dimension of moduli space does not
jump in this case. Then naturally
arises the question about the relation between (\ref{15}) and (\ref{17}).
However we want to proceed further with the formula
(\ref{17}).

Then we want to use the torus action method invented by Kontsevich,
and we couple gravity to the topological
sigma model. Roughly speaking, we add to the moduli space
``puncture'' degrees of freedom. So for k-point
correlation function, dimension of moduli space
(we denote it as ${\cal M}_{0,n,k}^{CP^{N-1}}$) increases by
$k-3$. $-3$ corresponds to deviding by automorphism of
$CP^{1}$, i.e.$SL(2,C)$
which is induced by $c$-ghost
zero-modes. And topological selection rule (\ref{sel}) is changed
into
\begin{eqnarray}
 N-2+k-3= \sum_{i=1}^{k} j_{i}\nonumber\\
\Longleftrightarrow \quad N-5 =\sum_{i=1}^{k}(j_{i}-1)
\label{18}
\end{eqnarray}
${\cal M}_{0,n,k}^{CP^{N-1}}$ can be roughly represented as follows,
\begin{equation}
{\cal M}_{0,n,k}^{CP^{N-1}} \simeq \{(z_{1},z_{2},\cdots,z_{k}),f \}/SL(2,C)
\qquad f \in {\cal M}_{0,n}^{CP^{N-1}}
\end{equation}
where $u \in SL(2,C)$ acts
\begin{equation}
u \circ \{(z_{1},z_{2},\cdots,z_{k}),f\} =\{(u(z_{1}),\cdots,u(z_{k})),
(u^{-1})^{*}\circ f \}
\end{equation}
This action of $SL(2,C)$ is compatible with the ``evaluation map'',
\begin{eqnarray}
\phi_{i} : {\cal M}_{0,n,k}^{CP^{N-1}} \mapsto CP^{N-1}\nonumber\\
\{(z_{1},z_{2},\cdots,z_{k}),f\} \mapsto f(z_{i})
\label{19}
\end{eqnarray}
because $(u^{-1})^{*}f(u(z_{i})) = f(z_{i})$.

Then the integral representation of amplitudes (\ref{17}) turns into
\begin{eqnarray}
\langle {\cal O}_{e^{j_{1}}}(z_{1}){\cal O}_{e^{j_{2}}}(z_{2}) \cdots
{\cal O}_{e^{j_{k}}}(z_{k}) \rangle _{n,alt,grav.} \nonumber\\
= \int_{{\cal M}_{0,n,k}^{CP^{N-1}}} c_{n}(M_{N})
\wedge \phi^{*}_{1}(c_{1}^{j_1}(H)) \wedge \cdots \wedge \phi^{*}_{k}
(c_{1}^{j_{k}}(H))
\label{20}
\end{eqnarray}
where we used the fact that $e$ corresponds to the first chern class of
hyperplane bundle $H$.
Then we have to find find the realization of $c_{n}(M_{N})$. We can roughly
do it as follows.
First consider the coordinate representation of ${\cal M}_{0,n}^{CP^{N-1}}$,
\begin{equation}
f:(s,t) \mapsto (\sum_{i=0}^{n}a_{1}^{i}s^{n-i}t^{i},
\cdots,
\sum_{i=0}^{n}a_{N}^{i}s^{n-i}t^{i} )
\label{21}
\end{equation}
where $(a_{i}^{j})$'s are the coordinates of ${\cal M}_{0,n}^{CP^{N-1}}$.
Then the condition imposed by
$c_{n}(M_{N})$ is equal to
\begin{eqnarray}
&&f(CP^{1}) \in M_{N} \quad \mbox{for all} (s,t)\nonumber\\
&\Longleftrightarrow& (\sum_{i=0}^{n}a_{1}^{i}s^{n-i}t^{i})^{N}+
\cdots +(\sum_{i=0}^{n}a_{N}^{i}s^{n-i}t^{i})^{N}=0
\quad \mbox{for all}(s,t)\nonumber\\
&\Longleftrightarrow& f^{m}(a_{j}^{i}) = 0\qquad (m=0,1,\cdots ,Nn)
\label{22}
\end{eqnarray}
where $f^{m}(a_{j}^{i})$'s are the coefficient polynomial of $s^{m}t^{Nn-m}$
of the l.h.s of the second line
of (\ref{22}).This imposes $Nn+1$ condition on ${\cal M}_{0,n}^{CP^{N-1}}$.
We can describe this
condition  mathematically in terms of gravitatinal moduli space
${\cal M}_{0,n,k}^{CP^{N-1}}$.
Let $\pi_{j}$ be a forgetfull map $\pi_{j}:{\cal M}_{0,n,j}^{CP^{N-1}}
\rightarrow {\cal M}_{0,n,j-1}^{CP^{N-1}}$.
Then for $j=1$, the fiber of $\pi_{1}$ is $CP^{1}$. And consider the sheaf
$\phi_{1}^{*}(N H)$ on
${\cal M}_{0,n,1}^{CP^{N-1}}$ where $N H$ corresponds to defining
polynomial of $M_{N}$ and
$H^{0}({\cal M}_{0,n,1}^{CP^{N-1}},\phi_{1}^{*}(N H))$ to the second
line of (\ref{22}) modulo
$SL(2,C)$ equivalence. Then direct limit sheaf
$R^{0}_{\pi_{1}}(\phi^{*}_{1}(N H))$(we denote it as
${\cal E}_{Nn+1}$).It locally equals $H^{0}(CP^{1},f^{*}{\cal O}(N H))$ has
rank
$(Nn+1)$.We can translate the
operation in going from the second line of (\ref{22}) to the third one into
the evaluation of the zero
locus of the section of ${\cal E}_{Nn+1}$.
 Considering the map,
\begin{equation}
\tilde{\pi}_{k}:= \pi_{1} \circ \pi_{2} \circ \cdots \circ \pi_{k} :
{\cal M}_{0,n,k}^{CP^{N-1}} \mapsto {\cal M}_{0,n,0}^{CP^{N-1}}
\label{23}
\end{equation}
We have
\begin{equation}
c(M_{N})= c_{T}(\tilde{\pi}_{k}^{*}({\cal E}_{Nn+1}))
\label{24}
\end{equation}
Finally the representation (\ref{20}) turns into
\begin{eqnarray}
\langle {\cal O}_{e^{j_{1}}}(z_{1}){\cal O}_{e^{j_{2}}}(z_{2}) \cdots
{\cal O}_{e^{j_{k}}}(z_{k}) \rangle _{n,alt,grav.} \nonumber\\
= \int_{{\cal M}_{0,n,k}^{CP^{N-1}}}
c_{T}(\tilde{\pi}_{k}^{*}({\cal E}_{Nn+1}))
\wedge \phi^{*}_{1}(c_{1}^{j_1}(H)) \wedge \cdots
\wedge \phi^{*}_{k}(c_{1}^{j_{k}}(H))
\label{25}
\end{eqnarray}
 Then we can calculate the correlation functions using
the torus action method.
\section{Review of the Torus Action Method}
\subsection{Introduction of the Torus Action and the Bott Residue Formula}
Torus action method is the strategy to use the Bott residue formula
\cite{bott} which reduces
the integral of Chern classes of vector bundle on $X$ to the one
on $X_{f}$ of the fixed point set of the
torus action flow on $X$ to the case where
$X$ is ${\cal M}_{0,n,k}^{CP^{N-1}}$.

First, let us introduce the torus action flow on $CP^{N-1}$,
\begin{eqnarray}
T_{t}: CP^{N-1}& \mapsto& CP^{N-1}\nonumber\\
   (X_{1},X_{2},\cdots,X_{N})& \mapsto &
(e^{\lambda_{1}t}X_{1},e^{\lambda_{2}t}X_{2},\cdots,e^{\lambda_{N}t}X_{N})
\nonumber\\
&& (t \in C^{\times})
\label{26}
\end{eqnarray}
 where $\lambda_{i} \in C$ is the character of the flow.Then (\ref{26})
induce the flow on ${\cal M}_{0,n,k}^{CP^{N-1}}$
from the compatibility with the evaluation map.
\begin{eqnarray}
 &&\phi_{i}(T_{t}((z_{1},z_{2},\cdots ,z_{k},f)/\sim))\nonumber\\
 &:=& T_{t} \circ \phi_{i}((z_{1},z_{2},\cdots ,z_{k},f)/\sim)\nonumber\\
 & =& T_{t} \circ f(z_{i})
\label{27}
\end{eqnarray}
  Next, we introduce the Bott residue formula. For simplicity,
we use $X$ for ${\cal M}_{0,n,k}^{CP^{N-1}}$.Let
${\cal E}_{1},\cdots ,{\cal E}_{m}$ be a holomorphic vector bundle on $X$,
and $X_{f}$ be the fixed point set of
$X$ under the flow $(\ref{27})$. We can decompose $X_{f}$ as the sum
of the connected components $X_{\gamma}$.
\begin{equation}
X_{f} = \bigcup_{\gamma} X_{\gamma}
\label{28}
\end{equation}
Then consider ${\cal E}_{i}|_{X_{\gamma}}$ and the normal bundle
${\cal N}_{\gamma} \simeq T^{\prime}X|_{X_{\gamma}}/T^{\prime}X_{\gamma}$
and decompose them into the eigen vector
bundle under the torus action $T_{t}$,i.e.,
\begin{eqnarray}
 {\cal E}_{i}|_{X_{\gamma}} &\cong& \bigoplus_{j=1}^{m_{{\cal E}_{i}}}
{\cal E}_{i,j}^{\gamma,f_{i,j}(\lambda_{*})}\nonumber\\
 {\cal N}_{\gamma} &\cong& \bigoplus_{j=1}^{m_{{\cal N}}}
{\cal N}_{j}^{\gamma,g_{j}(\lambda_{*})}
\label{29}
\end{eqnarray}
where
\begin{eqnarray}
T_{t}({\cal E}_{i,j}^{\gamma,f_{i,j}(\lambda_{*})})& = &
e^{f_{i,j}(\lambda_{*})t}{\cal E}_{i,j}^{\gamma,f_{i,j}(\lambda_{*})}
\nonumber\\
T_{t}({\cal N}_{j}^{\gamma,g_{j}(\lambda_{*})})& = & e^{g_{j}(\lambda_{*})t}
{\cal N}_{j}^{\gamma,g_{j}(\lambda_{*})}
\label{31}
\end{eqnarray}
and we set
\begin{eqnarray}
rk({\cal E}_{i,j}^{\gamma,f_{i,j}(\lambda_{*})}) = r_{\cal E}(i,j)\nonumber\\
rk({\cal N}_{j}^{\gamma,g_{j}(\lambda_{*})}) = r_{\cal N}(j)
\label{32}
\end{eqnarray}
 We can represent the total Chern class of
${\cal E}_{i,j}^{\gamma,f_{i,j}
(\lambda_{*})}$ and ${\cal N}_{j}^{\gamma,g_{j}(\lambda_{*})}$
as the product of first Chern class of formal line bundles as follows.
\begin{eqnarray}
c({\cal E}_{i,j}^{\gamma,f_{i,j}(\lambda_{*})}|_{X_{\gamma}} ) &=&
\prod_{j=1}^{m_{{\cal E}_{i}}}\prod_{k=1}^{r_{\cal E}(i,j)}
(1+t \cdot e_{i,j,k}^{\gamma,f_{i,j}(\lambda_{*})}) \nonumber\\
c({\cal N}_{\gamma}) &=&
\prod_{j=1}^{m_{{\cal N}}}\prod_{k=1}^{r_{\cal N}(j)}
(1+t \cdot n_{j,k}^{\gamma,g_{j}(\lambda_{*})})
\label{33}
\end{eqnarray}
Top Chern classes are given as the coefficient form of $t^{k}$ of highest
degree.

With these preparations, we introduce the Bott residue formula.

\begin{eqnarray}
\lefteqn{\int_{X} \prod_{i} c_{T}^{\alpha_{i}}({\cal E}_{i}) =} \nonumber\\
&&\sum_{\gamma} \int_{X_{\gamma}}
\frac{\prod_{i}\prod_{j=1}^{m_{{\cal E}_{i}}}
\prod_{k=1}^{r_{{\cal E}}(i,j)}(e_{i,j,k}^{\gamma,f_{i,j}(\lambda_{*})}
+f_{i,j}(\lambda_{*}))^{\alpha_{i}}}{\prod_{j=1}^{m_{\cal N}}
\prod_{k=1}^{r_{\cal N}(i)}
(n_{i,j}^{\gamma,g_{j}(\lambda_{*})}+g_{j}(\lambda_{*}))}
\label{34}
\end{eqnarray}
\subsection{Construction of Fixed Point Set}
Fixed points of $CP^{N-1}$ under $T_{t}$ are given by the projective
equivalence
\begin{equation}
p_{i} := (0,0,\cdots,0,\overbrace{1}^{i},0,\cdots,0)
\label{35}
\end{equation}
Then, we can find the fundamental maps $l_{i,j}^{d}$ from $CP^{1} \mapsto
CP^{N-1}$ which remain fixed under
$T_{t}$ as the degree d maps which connect $p_{i}$ and $p_{j}$.
\begin{equation}
l_{i,j}^{d}: (s,t)
\mapsto (0,\cdots,0,\overbrace{s^{d}}^{i},
0,\cdots,0,\overbrace{t^{d}}^{j},0,\cdots,0)
\label{36}
\end{equation}
 Of course $l_{i,j}^{d}$ is kept fixed under $SL(2,C)$ equivalence.
But now that we have coupled gravity with
the theory, we have to consider the boundary components of moduli space of
$CP^{1}$, i.e., stable curves. Stable
curve ${\cal C}$ with $k$-punctures is constructed with the set of
$CP^{1}$'s $\{C_{\alpha}\}$ with punctures
assigned on them and additional punctures of double singularity
which connect two components of $C_{\alpha}$'s.
Then we can translate the condition into the condition that the
genus of stable curve is zero
into imposing its arithmetic genus
to be zero.In geometrical language, if we represent $C_{\alpha}$ as a
line and define a figure with lines
which intersect at sigular punctures, this is equivalent to the non-existence
of closed loops in it. This addition
makes us to introduce stable maps which map stable curves to $CP^{N-1}$.

With these considerations, we can label the connected components of
the fixed
point set
${\cal M}_{0,n,k,f}^{CP^{N-1}}$ with a tree graph $\Gamma$ with the
following
structure. We denote them by
${\cal M}_{0,n,k}^{CP^{N-1}}(\Gamma)$.  The rules of
correspondences are,
\begin{itemize}
\item[1)] The vertices $v \in Vert(\Gamma)$ correspond to the connected
component
$C_{v}$ of\\
 $f^{-1}(p_{1},\cdots,p_{N})$.
$C_{v}$ can be a sum of connected irreducible components of ${\cal C}$
or be a point.

\item[2)] The edges ${\alpha} \in Edge(\Gamma)$ correspond to the irreducible
component $C_{\alpha}$mapped to $l_{i,j}^{d}$.
\end{itemize}
Then we have to add the additional structures to $\Gamma$,
\begin{itemize}
\item[1)] We label each $v \in Vert(\Gamma)$ by $f_{v} \in \{1,2,\cdots ,N\}$
which is defined by $p_{f_{v}} = f(C_{v})$.

\item[2)] The $k$-punctures  are distributed among the vertices
$v \in Vert(\Gamma)$.
We represent this distribution by
$S_{v} \in \{1,2,\cdots,k\}$.

\item[3)] We attach degree $d_{\alpha}$ to each $\alpha \in
Edge(\Gamma)$
defined by the degree of $l_{i,j}^{d}$.
\end{itemize}
We have to set punctures on the vertices $Vert(\Gamma)$ because
if we put punctures
on $C^{\alpha}$, they move with the flow
$T_{t}$, which contradicts with the assumption of fixed point sets.
Then we can construct
${\cal M}_{0,n,k}^{CP^{N-1}}(\Gamma)$ under conditions that emerges from
the above three structures,
\begin{itemize}
\item[1)] If ${\alpha} \in Edge(\Gamma)$ connects $v,u \in Vert(\Gamma)$,
$f_{u} \neq f_{v}$.

\item[2)] $\{1,2,\cdots,k\} = \coprod_{v \in Vert(\Gamma)} S_{v}$.

\item[3)] $\sum_{\alpha \in Edge(\Gamma)} d_{\alpha} = n$
\end{itemize}
Then we have
\begin{equation}
{\cal M}_{0,n,k}^{CP^{N-1}}(\Gamma) \cong \prod_{v \in Vert(\Gamma)}
({\cal M}_{0,S_{v}})/(Aut(\Gamma))
\label{38}
\end{equation}
where ${\cal M}_{0,S_{v}}$ is the moduli space of complex structure of
$CP^{1}$ with $S_{v}$ punctures. It represents
the gravitational degree of freedom of $C_{v}$.According to Kontsevich,
division by $Aut(\Gamma)$ reflects the orbispace
structure of ${\cal M}_{0,n,k}^{CP^{N-1}}$. It may reflect
the multiplicity of the degeneration of stable maps.
\subsection{Determination of the contribution from Normal and Vector bundles}
\subsubsection{Contributions from ${\cal N}_{{\cal M}(\Gamma)}^{abs}$}
With these preparations, we determine the contribution from
${\cal M}_{0,n,k}^{CP^{N-1}}$ (in the following discussion
we abbriviate the notation as ${\cal M}(\Gamma))$ to (\ref{34}).

First, we calculate the contribution from ${\cal N}_{{\cal M}(\Gamma)}$.
Following Kontsevich,we will use the expression
of vector bundles as the K-group [ ],which translates sum and quotient
operations into addition and subtraction.
Then we have
\begin{equation}
[{\cal N}_{{\cal M}(\Gamma)}] = [T^{\prime}{\cal M}|_{{\cal M}(\Gamma)}]
-[T^{\prime}{\cal M}(\Gamma)]
\label{39}
\end{equation}
 If we set $${\cal C}= \tilde{\bigcup}_{\alpha} C_{\alpha}$$
(where $$\tilde{\bigcup}_{\alpha}$$ means a sum with
double-singularity gluing operation.),
$[T^{\prime}{\cal M}|_{{\cal M}(\Gamma)}]$ consists of the following degrees
of freedom,
\begin{itemize}
\item[1)] Moving $f({\cal C})$ in $CP^{N-1}$.

\item[2)] Resolution of singularities of ${\cal C}$, i.e., from $xy =0$ to
$xy = \epsilon$.

\item[3)] Moving puncture degrees of freedom.
\end{itemize}
And we have
\begin{eqnarray}
[T^{\prime}{\cal M}|_{{\cal M}(\Gamma)}] &=&
[H^{0}({\cal C},f^{*}(T^{\prime}CP^{N-1}))]\nonumber\\
        &+& \sum_{{z \in C_{\alpha} \cap C_{\beta}}
\atop {{\alpha}\ne{\beta}}}
[T^{\prime}_{z}C_{\alpha}\otimes T^{\prime}_{z}C_{\beta}]\nonumber\\
&+& \sum_{{z \in C_{\alpha} \cap C_{\beta}} \atop {{\alpha}\ne {\beta}}}
[T^{\prime}_{z}C_{\alpha}]+ [T^{\prime}_{z}C_{\beta}]\nonumber\\
&+&\sum_{z_{i},i \in \{1,\cdots,k\}}[T^{\prime}_{z_{i}}{\cal C}]-
\sum_{\alpha}[H^{0}(C^{\alpha},T^{\prime}C^{\alpha})]
\label{40}
\end{eqnarray}
 The last term of (\ref{40}) corresponds to devision by $SL(2,C)$ of
each component $C_{\alpha}$.
{}From (\ref{38}) ${\cal M}(\Gamma)$ has continuous degrees of freedom which
come only from $C_{\alpha}$ mapped to a
point, we have
\begin{eqnarray}
[T^{\prime}{\cal M}(\Gamma)]& =& \sum_{{z \in C_{\alpha} \cap C_{\beta}}
\atop{\alpha \ne \beta\;\alpha,\beta \notin Edge(\Gamma)}}
 [T^{\prime}_{z}C_{\alpha}\otimes T^{\prime}_{z}C_{\beta}] \nonumber\\
 &+&  \sum_{{z \in C_{\alpha} \cap C_{\beta}} \atop {\alpha \ne \beta,
\alpha \notin Edge(\Gamma)}}
[T^{\prime}_{z}C_{\alpha}]\nonumber\\
&+& \sum_{z_{i},i \in \{1,2,\cdots,k\}} [T^{\prime}_{z_{i}}{\cal C}]\nonumber\\
&-& \sum_{\alpha\notin Edge(\Gamma)}[H^{0}(C^{\alpha},T^{\prime}C^{\alpha})]
\label{41}
\end{eqnarray}
where we used the fact that all the punctures lie in the component mapped to
a point .

{}From (\ref{40}) and (\ref{41}), we have
\begin{equation}
[{\cal N}_{{\cal M}(\Gamma)}] = [H^{0}({\cal C},f^{*}(T^{\prime}CP^{N-1}))]
+ [{\cal N}_{{\cal M}(\Gamma)}^{abs}]
\label{42}
\end{equation}
where
\begin{eqnarray}
[{\cal N}_{{\cal M}(\Gamma)}^{abs}] &:=
&\sum_{{z \in C_{\alpha} \cap C_{\beta}}\atop{\alpha \ne \beta\;
\alpha,\beta \in Edge(\Gamma)}}
 [T^{\prime}_{z}C_{\alpha}\otimes T^{\prime}_{z}C_{\beta}] \\
\label{43}
 &+&  \sum_{{z \in C_{\alpha} \cap C_{\beta}}
\atop {\alpha \in Edge(\Gamma),\beta\notin Edge(\Gamma)}}
[T^{\prime}_{z}C_{\alpha}\otimes T^{\prime}_{z}C_{\beta}]\\
\label{44}
&+& \sum_{{z \in C_{\alpha} \cap C_{\beta}}\atop{\alpha \ne \beta
\;\alpha\in Edge(\Gamma)}} [T^{\prime}_{z}C_{\alpha}]
- \sum_{\alpha\in Edge(\Gamma)}
[H^{0}(C^{\alpha},T^{\prime}C^{\alpha})]
\label{45}
\end{eqnarray}
Then we determine the contribution from the first term of
(\ref{42}),and
(\ref{43}),(\ref{44}),and (\ref{45}).

First, consider the contribution from (\ref{43}).
Since $\alpha,\beta \in Edge(\Gamma)$,
 $T_{z}C_{\alpha}$ and
 $T_{z}C_{\beta}$'s
are trivial as the line bundle on ${\cal M}(\Gamma)$.
Let $C_{\alpha}$ and $C_{\beta}$ be mapped to $l_{i,j}^{d_{\alpha}}$ and
$l_{i,l}^{d_{\beta}}$.
\begin{eqnarray}
C_{\alpha} : (z_{1},z_{2}) \mapsto
(0,\cdots,0,\overbrace{z_{1}^{d_{\alpha}}}^{i},0,\cdots,0,
\overbrace{z_{2}^{d_{\alpha}}}^{j},0,\cdots,0)
\\
\label{46}
C_{\beta} : (w_{1},w_{2}) \mapsto (0,\cdots,0,
\overbrace{w_{1}^{d_{\beta}}}^{i}
,0,\cdots,0,\overbrace{w_{2}^{d_{\beta}}}^{l},0,\cdots,0)
\\
\label{47}
\end{eqnarray}
Local coordinate around $z \in C_{\alpha} \cap C_{\beta}$ on
$C_{\alpha}$ and $C_{\beta}$ are $\frac{z_{2}}{z_{1}}$ and
$\frac{w_{2}}{w_{1}}$
,and we have
\begin{equation}
T^{\prime}_{z}C_{\alpha}\otimes T^{\prime}_{z}C_{\beta}
\cong \frac{d}{d(\frac{z_{2}}{z_{1}})}\otimes\frac{d}{d(\frac{w_{2}}{w_{1}})}
\label{48}
\end{equation}
Definition of torus action (\ref{27}) leads us to
\begin{eqnarray}
z_{1} \mapsto z_{1}e^{\frac{\lambda_{i}}{d_{\alpha}}t}&z_{2}
\mapsto z_{2}e^{\frac{\lambda_{j}}{d_{\alpha}}t}\nonumber\\
w_{1} \mapsto w_{1}e^{\frac{\lambda_{i}}{d_{\beta}}t}&w_{2}
\mapsto w_{2}e^{\frac{\lambda_{l}}{d_{\beta}}t}
\label{49}
\end{eqnarray}
and
\begin{equation}
T^{\prime}_{z}C_{\alpha}\otimes T^{\prime}_{z}C_{\beta}
\mapsto e^{(\frac{\lambda_{i}-\lambda_{j}}{d_{\alpha}}+
\frac{\lambda_{i}-\lambda_{l}}{d_{\beta}})t}
T^{\prime}_{z}C_{\alpha}\otimes T^{\prime}_{z}C_{\beta}
\label{50}
\end{equation}
The result is,
\begin{equation}
(\mbox {Contribution from (\ref{43}) to (\ref{34})}) =
\prod_{{C_{\alpha}\cap C_{\beta}\neq\emptyset}\atop
{\alpha\neq\beta\;\alpha,\beta\in Edge(\Gamma)}}
\frac{1}{\frac{\lambda_{i}-\lambda_{j}}{d_{\alpha}}+
\frac{\lambda_{i}-\lambda_{l}}{d_{\beta}}}
\label{51}
\end{equation}
Again following Kontsevich, we introduce the notation ``Flag''
$F=(v,\alpha)$ which represents edge $\alpha$ with a direction
specified by the source vertex $v$. We define
\begin{equation}
w_{F}:= \frac{\lambda_{f_{v}}-\lambda_{f_{u}}}{d_{\alpha}}
\label{52}
\end{equation}
Then the r.h.s of (\ref{51}) can be rewritten as follows.
\begin{equation}
\prod_{{v\in Vert(\Gamma)}\atop{val(v)=2}\quad{{}^{\sharp}S_{v}}=0}
\frac{1}{w_{F_{1}(v)}+w_{F_{2}(v)}}
\label{53}
\end{equation}
where $val(v)$ represents the valency of $v$ and $F_{1}(v)$ and $F_{2}(v)$
are the flags whose sources are $v$. Note that in this case
$f^{-1}(v)$ is a point.

Next we consider the contributions from (\ref{44}).
Again from (\ref{41}), $T_{z}C_{\alpha}$ is trivial as the line bundle on
${\cal M}(\Gamma)$ but has an eigenvalue $w_{F}$ as in
the derivation of (\ref{50}).On the other hand, $T^{\prime}_{z}C_{\beta}$
has trivial torus action (because $C_{\beta}$ is mapped to a point)
but non trivial line bundle on ${\cal M}(\Gamma)$.And if
${}^{\sharp}(\mbox {punctures on} C_{v})\geq 3$,${\cal M}_{0,S_{v}}$
is well-defined and we have
\begin{eqnarray}
(\mbox{Contribution from (\ref{44}) to (\ref{34})})
&=& \prod_{v \in Vert(\Gamma)}(\int_{{\cal M}_{0,val(v)+{}^{\sharp}S_{v}}}
\prod_{{flags}\atop {F=(v,\alpha)}}
\frac{1}{w_{F}+c_{1}(T^{\prime}_{z_{F}}(C_{v}))})\nonumber\\
&&(val(v)+{}^{\sharp}S_{v} \geq 3)
\label{54}
\end{eqnarray}
where $z_{F}$ represents the gluing point of $C_{v}$ and $F$.
We can evaluate the r.h.s of (\ref{54}) by expanding in terms of
$\frac{1}{w_{F}}$ and using the fact that
$c_{1}(T^{\prime}_{z_{F}}C_{v}) = - c_{1}(T^{\prime,*}_{z_{F}}C_{v})$.
Expansion coefficients are intersection numbers of
Mumford-Morita class on the $CP^{1}$-moduli space ,which is identified as the
correlation function of gravitational
descendants by Witten \cite{witten2}.Continuing the calculation,we have
\begin{equation}
(\mbox{r.h.s of (\ref{54})})= \prod_{v \in Vert(\Gamma)} (\sum_{{d_{1},\cdots,
d_{val(v)}\geq 0}\atop{\sum d_{i}= val(v)+{}^{\sharp}S_{v}-3}}
\prod_{{flags}\atop{F=(v,\alpha)}}w_{F}^{-d_{i}-1}\langle
\sigma_{d_{1}}\cdots\sigma_{d_{val{v}}}
\overbrace{P\cdots P}^{{}^{\sharp}S_{v}\mbox{times}}\rangle)
\label{55}
\end{equation}
 $\langle \sigma_{d_{1}}\cdots\sigma_{d_{val{v}}}\overbrace{P\cdots P}^
{{}^{\sharp}S_{v}\mbox{times}}\rangle$ is calculated in \cite{dij},
\begin{equation}
\langle \sigma_{d_{1}}\cdots\sigma_{d_{val{v}}}\overbrace{P\cdots P}^
{{}^{\sharp}S_{v}\mbox{times}}\rangle =
\frac{(val(v)+{}^{\sharp}S_{v}-3)!}{d_{1}!\cdots d_{val(v)}!}
\label{56}
\end{equation}
Combining (\ref{54}),(\ref{55}) and (\ref{56}), we have
\begin{eqnarray}
(\mbox{Contribution from (\ref{44}) to (\ref{34})}) &=&
\prod_{v \in Vert(\Gamma)}\prod_{{flags}\atop{F=(v,\alpha)}}
w_{F}^{-1}(\sum_{{flags}\atop {F=(v,\alpha)}}w_{F}^{-1})^{val(v)+
{}^{\sharp}S_{v}-3}\nonumber\\
&&(val(v)+{}^{\sharp}S_{v} \geq 3)
\label{57}
\end{eqnarray}
Then we consider (\ref{45}). Contributions of the first terms are, as before
\begin{equation}
\prod_{{C_{\alpha}\cap C_{\beta}}\atop
{\alpha\neq\beta\;\alpha\in Edge(\Gamma)}}\frac{1}{w_{F_{i}(\alpha)}}
\label{58}
\end{equation}
where $F_{i}(\alpha)$'s are two flags having $\alpha$ as their edges.

The second terms which represents the automorphism group degrees of freedom
of edge components can be expressed by
the tangent bundles on the inverse images of two vertices of the edges and
scaling transformation degree of freedom
fixing the punctures (We denote it as $[0]$). In terms of the
K-group, we have
\begin{eqnarray}
-\sum_{\alpha \in Edge(\Gamma)}[H^{0}(C_{\alpha},T^{\prime}C_{\alpha})]
\nonumber\\
= - \sum_{\alpha \in Edge(\Gamma)}([T^{\prime}_{z_{1}(\alpha)}C_{\alpha}]
+[0]+[T^{\prime}_{z_{2}(\alpha)}C_{\alpha}])
\label{59}
\end{eqnarray}
And contributions to (\ref{34}) are
\begin{equation}
\prod_{\alpha \in Edge(\Gamma)}w_{F_{1}}(\alpha)\cdot
w_{F_{2}}(\alpha)\cdot C([0])
\label{60}
\end{equation}
where $C([0])$ represents the factor from $[0]$.
Multiplying (\ref{58}) and (\ref{60}), what remains except for $C([0])$
is the products of $w_{F}$'s whose edges have only one double singularity.
In other words, the corresponding $F= (v,\alpha)$
has $val(v) = 1$ and $f^{-1}(v)$ is a point. We have
\begin{equation}
(\mbox{Contributions from (\ref{45})}) =
\prod_{{v\in Vert(\Gamma)}\atop{{val(v)=1}\quad{{}^{\sharp}S_{v}=0}}}
\prod_{{flags}\atop {F=(v,\alpha)}}w_{F}\prod_{\alpha\in Edge(\Gamma)}C([0])
\label{61}
\end{equation}
After all, from (\ref{53}),(\ref{57}) and (\ref{61}), we put all the factors
from $[{\cal N}_{{\cal M}(\Gamma)}^{abs}]$
into the form,
\begin{equation}
\prod_{v\in Vert(\Gamma)}\prod_{{flags}\atop {F=(v,\alpha)}}w_{F}^{-1}
(\sum_{{flags}\atop{F=(v,\alpha)}}w_{F}^{-1})^{val(v)
+{}^{\sharp}S_{v}-3}\prod_{\alpha\in Edge(\Gamma)}C([0])
\label{62}
\end{equation}
\subsubsection{Determination of the contributions from
$[H^{0}({\cal C},f^{*}(T^{\prime}CP^{N-1}))]$}
Since $$f({\cal C})= \tilde{\bigcup}_{\alpha\in Edge(\Gamma)}f(C_{\alpha})$$,
 we can construct
 $[H^{0}({\cal C},f^{*}(T^{\prime}CP^{N-1}))]$ by gluing
$$\bigoplus_{\alpha\in Edge(\Gamma)}
[H^{0}(C_{\alpha},f^{*}(T^{\prime}CP^{N-1}))]$$
at $p_{f_{v}}$.This process can be described using exact sequences,
\begin{eqnarray}
0 \mapsto H^{0}({\cal C},f^{*}(T^{\prime}CP^{N-1}))\mapsto\nonumber\\
\bigoplus_{\alpha\in Edge(\Gamma)}H^{0}(C_{\alpha},f^{*}(T^{\prime}CP^{N-1}))
\mapsto
\bigoplus_{v \in Vert(\Gamma)}C^{val(v)-1}\otimes
T^{\prime}_{p_{f_{v}}}CP^{N-1} \mapsto 0
\label{63}
\end{eqnarray}
This contribution is then given as the contribution from the second term
divided by the one from from the third term. As the independent
basis of $H^{0}(C_{\alpha},f^{*}(T^{\prime}CP^{N-1}))$ describing the
deformation of $f(C_{\alpha})$ in $CP^{N-1}$ where $C_{\alpha}$
is
\begin{equation}
C_{\alpha}:\quad (z_{1},z_{2})\mapsto (0,\cdots,0,
\overbrace{z_{1}^{d_{\alpha}}}^{i},0,\cdots,0,
\overbrace{z_{2}^{d_{\alpha}}}^{j},
0,\cdots,0)
\label{64}
\end{equation}
, we have
\begin{eqnarray}
(0,\cdots,0,\overbrace{z_{1}^{d_{\alpha}}+
\epsilon z_{1}^{m}z_{2}^{d_{\alpha}-m}}^{i},
0,\cdots,0,\overbrace{z_{2}^{d_{\alpha}}}^{j},
0,\cdots,0)\\
\label{65}
(0,\cdots,0,\overbrace{z_{1}^{d_{\alpha}}}^{i},0,\cdots,0,
\overbrace{z_{2}^{d_{\alpha}}+
\epsilon z_{1}^{m}z_{2}^{d_{\alpha}-m}}^{j},
0,\cdots,0)\\
\label{66}
(0,\cdots,0,\overbrace{\epsilon z_{1}^{m}z_{2}^{d_{\alpha}-m}}^{k},0,\cdots,
0,\overbrace{z_{1}^{d_{\alpha}}}^{i},0,\cdots,0,
\overbrace{z_{2}^{d_{\alpha}}}^{j},0,\cdots,0)
\label{67}
\end{eqnarray}
We can write these basis in more sophisticated form,
\begin{eqnarray}
a)\qquad &(\frac{z_{1}}{z_{2}})^{m}X_{i}\frac{\partial}{\partial X_{i}}&
\qquad(-d_{\alpha}\leq m \leq d_{\alpha})\\
\label{68}
b)\qquad &(\frac{z_{1}}{z_{2}})^{m}X_{j}\frac{\partial}{\partial X_{k}}&
\qquad(0\leq m \leq d_{\alpha})\qquad k\neq i,j
\label{69}
\end{eqnarray}
This expression directly leads us to
\begin{eqnarray}
(\mbox{Contribution from (\ref{68})})& =
&\frac{1}{m\cdot w_{F}}\qquad (m\neq 0)\\
\label{70}
&&\frac{1}{C([0])}\qquad (m=0)\\
\label{70b}
(\mbox{Contribution from (\ref{69})})& =
&\frac{1}{m\cdot w_{F}+\lambda_{j}-\lambda_{k}}
\label{71}
\end{eqnarray}
And from $T^{\prime}_{p_{f_{v}}}CP^{N-1}\simeq
\oplus_{j\neq p_{f_{v}}}\frac{\partial}{\partial(\frac{X_{j}}{X_{f_{v}}})}$,
\begin{equation}
(\mbox{Contributions
from $C^{val(v)-1}\otimes T^{\prime}_{p_{f_{v}}}CP^{N-1}$})
=\prod_{v\in Vert(\Gamma)}
(\lambda_{f_{v}}-\lambda_{j})^{val(v)-1}
\label{72}
\end{equation}
Combining (\ref{70}),(\ref{70b}),(\ref{71}) and (\ref{72}), we have
\begin{eqnarray}
\prod_{flags\quad F=(v,\alpha)}(w_{F})^{-d_{\alpha}}
\prod_{{\alpha\in Edge(\Gamma)}\atop{(u,v): vertices\;of\:\alpha}}
\prod_{k\neq f_{u},f_{v}}\prod_{m=0}^{d_{\alpha}}
(\frac{m\lambda_{f_{u}}+
(d_{\alpha}-m)\lambda_{f_{v}}}{d_{\alpha}}-\lambda_{k})^{-1}
\nonumber\\
\prod_{\alpha\in Edge(\Gamma)}(d_{\alpha}!)^{-2}
(C([0]))^{-1}\prod_{v\in Vert(\Gamma)}(\prod_{j:j\neq f_{v}}
(\lambda_{f_{v}}-\lambda_{j}))^{val(v)-1}
\label{73}
\end{eqnarray}
\subsubsection{Factors from Vector Bundles ${\cal E}_{i}$}
First, we calculate the factors from ${\cal E}_{Nn+1}$. As we have
mentioned
in section 2, this fiber locally
corresponds to $[H^{0}({\cal C},f^{*}({\cal O}(N\cdot H)))]$. We can construct
it as in (\ref{62}), by the exact sequences,
\begin{eqnarray}
0 \mapsto H^{0}({\cal C},f^{*}({\cal O}(N H)))\mapsto \nonumber\\
\bigoplus_{\alpha\in Edge(\Gamma)}H^{0}(C_{\alpha},f^{*}({\cal O}
(N H)))
\mapsto \bigoplus_{v \in Vert(\Gamma)}C^{val(v)-1}\otimes {\cal O}_{p_{f_{v}}}
(N H) \mapsto 0
\label{74}
\end{eqnarray}
Then since the basis of $H^{0}(C_{\alpha},f^{*}({\cal O}(N\cdot H)))$
are given as
\begin{equation}
\{z_{1}^{N d_{\alpha}},z_{1}^{N d_{\alpha}-1}z_{2},\cdots,z_{2}^{Nd_{\alpha}}\}
\label{75}
\end{equation}
and the section of ${\cal O}_{p_{f_{v}}}(N H)$ is
$X_{f_{v}}^{N}$, we have
\begin{eqnarray}
(\mbox{Contributions from $c_{T}({\cal E}_{Nn+1})$}) &=&
\prod_{{\alpha\in Edge(\Gamma)}\atop{(v_{1},v_{2}):
vertices\; of\; \alpha}}
\prod_{a=0}^{Nd_{\alpha}}(\frac{a\lambda_{f_{v_{1}}}+(N-a)
\lambda_{f_{v_{2}}}}{d_{\alpha}})\nonumber\\
&&\prod_{v\in Vert(\Gamma)}
(N\lambda_{f_{v}})^{1-val(v)}
\label{76}
\end{eqnarray}
Next, we determine the factor from $\phi^{*}_{i}(c_{1}^{j_{i}}(H))$.
{}From the argument of $\S 3.2$,puncture $i$ lies on the
vertex $v(i)$ of $\Gamma$,and $\phi^{*}_{i}(c_{1}^{j_{i}}(H))$ reduces to
${\cal O}_{p_{f_{v(i)}}}(j_{i}H)$. This leads
to
\begin{equation}
(\mbox{Contributions from $\phi^{*}_{i}(c_{1}^{j_{i}}(H))$}) =
\lambda^{j_{i}}_{f_{v(i)}}
\label{77}
\end{equation}
\subsubsection{Local Appendix}
We have to divide the above factors by ${}^{\sharp}Aut(\Gamma)$
coming from (\ref{38}) and in practice, we have to multiply a factor
$\frac{1}{d_{\alpha}}$ for each edge $\alpha$. We cannot justify
the reason for
this factor at this stage.
\section{Some Explict Calculation of Amplitudes}
{}From the argument of section 2, we constructed representation of amplitudes
as an integral of forms on ${\cal M}_{0,n,k}^{CP^{N-1}}$, and the strategy
of section 3 enables us to compute them as a sum of amplitudes corresponding
to tree graphs. For some examples, we calculate $\langle {\cal O}_{e^{N-4}}
\rangle_{1}$,$\langle {\cal O}_{e^{N-4}}\rangle_{2}$ and
$\langle {\cal O}_{e^{N-4}}\rangle_{3}$. First, we write out tree
graphs that
contribute to the amplitudes up to degree 3. (See {\bf Fig.1}.)
In {\bf Fig.1}, we abbriviate the external insertion
of ``punctures''.
So in calculation,we have to add all the cases of external operator
insertions of $\langle {\cal O}_{e^{N-4}}\rangle $ to vertices.
 Note that
the two character numbers (for example ``i'') of neighboring vertices
never coincide with each other. Then direct application of
the argument of
section 3 leads to the following formula.

\begin{eqnarray}
\langle {\cal O}_{e^{N-4}}\rangle_{1} &=&
\frac{1}{2}\sum_{i\ne j}(\prod_{k\ne i,j}(\lambda_{i}-\lambda_{k})^{-1}
(\lambda_{j}-\lambda_{k})^{-1}\prod_{a=0}^{N}(a\lambda_{i}+(N-a)\lambda_{j})
(\frac{\lambda_{i}^{N-4}-\lambda_{j}^{N-4}}{w_{F_{1}}}))\nonumber\\
&&(\mbox{from (a)})\nonumber\\
\langle {\cal O}_{e^{N-4}}\rangle_{2} &=&
\frac{1}{2}\sum_{{i\ne j}\atop{j\ne k}}(\frac{1}{N\lambda_{j}}
(\frac{\lambda^{N-4}_{i}w_{F_{4}} +  \lambda^{N-4}_{k}w_{F_{1}}}
{w_{F_{2}}+w_{F_{3}}} + \lambda_{j}^{N-4})\nonumber\\
&&\frac{1}{w_{F_{1}}w_{F_{2}}w_{F_{3}}w_{F_{4}}}
\prod_{n\ne j}(\lambda_{j}-\lambda_{n})\nonumber\\
&&\prod_{m_{1}\ne i,j}(\lambda_{i}-\lambda_{m_{1}})^{-1}
(\lambda_{j}-\lambda_{m_{1}})^{-1}
\prod_{m_{2}\ne j,k}(\lambda_{j}-\lambda_{m_{2}})^{-1}
(\lambda_{k}-\lambda_{m_{2}})^{-1}\nonumber\\
&&\prod_{a_{1}=0}^{N}(a_{1}\lambda_{i}+(N-a_{1})\lambda_{j})
\prod_{a_{2}=0}^{N}(a_{2}\lambda_{j}+(N-a_{2})\lambda_{k})\nonumber\\
&&(\mbox{from (b)})\nonumber\\
&+&\frac{1}{4}\sum_{i\ne j}((\lambda_{i}^{N-4}-\lambda_{j}^{N-4})w_{F_{2}}
\frac{1}{w_{F_{1}}^{2}w_{F_{2}}^{2}}\nonumber\\
&&\prod_{k\ne i,j}(\lambda_{i}-\lambda_{k})^{-1}
(\lambda_{j}-\lambda_{k})^{-1}(\frac{\lambda_{i}+\lambda_{j}}{2}
-\lambda_{k})^{-1}\nonumber\\
&&\prod_{a=0}^{2N}(\frac{a\lambda_{j}+(2N-a)\lambda_{k}}{2}))\nonumber\\
&&(\mbox{from (c)})\nonumber\\
\langle {\cal O}_{e^{N-4}} \rangle_{3}&=&
\frac{1}{2}\sum_{{i\ne j}\atop{j\ne k,k\ne l}}((
\lambda_{i}^{N-4}\frac{1}{w_{F_{2}}+w_{F_{3}}}\frac{1}{w_{F_{4}}+w_{F_{5}}}
w_{F_{6}}\nonumber\\
&+&w_{F_{1}}\frac{\lambda_{j}^{N-4}}{w_{F_{2}}w_{F_{3}}}
\frac{1}{w_{F_{4}}+w_{F_{5}}}
w_{F_{6}}\nonumber\\
&+&w_{F_{1}}\frac{1}{w_{F_{2}}+w_{F_{3}}}
\frac{\lambda_{k}^{N-4}}{w_{F_{4}}w_{F_{5}}}
w_{F_{6}}\nonumber\\
&+&w_{F_{1}}\frac{1}{w_{F_{2}}+w_{F_{3}}}
\frac{1}{w_{F_{4}}+w_{F_{5}}}
\lambda_{l}^{N-4})\nonumber\\
&&\frac{1}{w_{F_{1}}w_{F_{2}}w_{F_{3}}w_{F_{4}}w_{F_{5}}w_{F_{6}}}
\frac{1}{(N\lambda_{j}N\lambda_{k})}\nonumber\\
&&\prod_{n_{1}\ne j}(\lambda_{j}-\lambda_{n_{1}})
\prod_{n_{2}\ne k}(\lambda_{k}-\lambda_{n_{2}})\nonumber\\
&&\prod_{m_{1}\ne i,j}(\lambda_{i}-\lambda_{m_{1}})^{-1}
(\lambda_{j}-\lambda_{m_{1}})^{-1}\nonumber\\
&&\prod_{m_{2}\ne j,k}(\lambda_{j}-\lambda_{m_{2}})^{-1}
(\lambda_{k}-\lambda_{m_{2}})^{-1}\nonumber\\
&&\prod_{m_{3}\ne k,l}(\lambda_{k}-\lambda_{m_{3}})^{-1}
(\lambda_{l}-\lambda_{m_{3}})^{-1}\nonumber\\
&&\prod_{a_{1}=0}^{N}(a_{1}\lambda_{i}+(N-a_{1})\lambda_{j})
\prod_{a_{2}=0}^{N}(a_{2}\lambda_{j}+(N-a_{2})\lambda_{k})\nonumber\\
&&\prod_{a_{3}=0}^{N}(a_{3}\lambda_{k}+(N-a_{3})\lambda_{l}))\nonumber\\
&&(\mbox{from (d)})\nonumber\\
&+&\frac{1}{2}\sum_{{i\ne j}\atop{j\ne k}}(\frac{1}{4}
(\lambda_{i}^{N-4}\frac{1}{w_{F_{2}}+w_{F_{3}}}w_{F_{4}}+
w_{F_{1}}\frac{\lambda_{j}^{N-4}}{w_{F_{2}}w_{F_{3}}}w_{F_{4}}+
w_{F_{1}}\frac{1}{w_{F_{2}}+w_{F_{3}}}\lambda_{k}^{N-4})\nonumber\\
&&\frac{1}{N\lambda_{j}}\frac{1}{w_{F_{1}}^{2}w_{F_{2}}^{2}}
\frac{1}{w_{F_{3}}w_{F_{4}}}
\prod_{n\ne j}(\lambda_{j}-\lambda_{n})\nonumber\\
&&\prod_{m_{1}\ne i,j}(\lambda_{i}-\lambda_{m_{1}})^{-1}
(\lambda_{j}-\lambda_{m_{1}})^{-1}(\frac{\lambda_{i}+\lambda_{j}}{2}
-\lambda_{m_{1}})^{-1}\nonumber\\
&&\prod_{m_{2}\ne j,k}(\lambda_{j}-\lambda_{m_{2}})^{-1}
(\lambda_{k}-\lambda_{m_{2}})^{-1}\nonumber\\%
&&\prod_{a_{1}=0}^{2N}(\frac{a_{1}\lambda_{j}+(2N-a_{1})\lambda_{k}}{2})
\prod_{a_{2}=0}^{N}(a_{2}\lambda_{j}+(N-a_{2})\lambda_{k}))\nonumber\\
&&(\mbox{from (e)})\nonumber\\
&+&\frac{1}{6}\sum_{i\ne j}(\frac{1}{36}(w_{F_{2}}(\lambda_{i}^{N-4}-
\lambda_{j}^{N-4})\frac{1}{w_{F_{1}}^{3}w_{F_{2}}^{3}}\nonumber\\
&&\prod_{m\ne i,j}(\lambda_{i}-\lambda_{m})^{-1}
(\frac{2\lambda_{i}+\lambda_{j}}{3}-\lambda_{m})^{-1}
(\frac{\lambda_{i}+2\lambda_{j}}{3}-\lambda_{m})^{-1}
(\lambda_{j}-\lambda_{m})^{-1}\nonumber\\
&&\prod_{a=0}^{2N}(\frac{a\lambda_{i}+(3N-a)\lambda_{j}}{3}))\nonumber\\
&&(\mbox{from (f)})\nonumber\\
&+&\frac{1}{6}\sum_{{i\ne j}\atop{i\ne k,i\ne l}}(
\lambda_{j}^{N-4}\frac{1}{w_{F_{1}}w_{F_{3}}w_{F_{5}}}w_{F_{4}}w_{F_{6}}+
\lambda_{l}^{N-4}\frac{1}{w_{F_{1}}w_{F_{3}}w_{F_{5}}}w_{F_{2}}w_{F_{6}}
\nonumber\\
&+&\lambda_{k}^{N-4}\frac{1}{w_{F_{1}}w_{F_{3}}w_{F_{5}}}w_{F_{2}}w_{F_{4}}
\nonumber\\
&+&w_{F_{2}}w_{F_{4}}w_{F_{6}}\lambda_{i}^{N-4}
\frac{1}{w_{F_{1}}w_{F_{3}}w_{F_{5}}}
(\frac{1}{w_{F_{1}}}+\frac{1}{w_{F_{3}}}+\frac{1}{w_{F_{5}}}))\nonumber\\
&&\frac{1}{(N\lambda_{i})^{2}}
\frac{1}{w_{F_{1}}w_{F_{2}}w_{F_{3}}w_{F_{4}}w_{F_{5}}w_{F_{6}}}\nonumber\\
&&\prod_{m_{1}\ne i,j}(\lambda_{i}-\lambda_{m_{1}})^{-1}
(\lambda_{j}-\lambda_{m_{1}})^{-1}\nonumber\\
&&\prod_{m_{2}\ne i,l}(\lambda_{i}-\lambda_{m_{2}})^{-1}
(\lambda_{l}-\lambda_{m_{2}})^{-1}\nonumber\\
&&\prod_{m_{3}\ne i,k}(\lambda_{i}-\lambda_{m_{3}})^{-1}
(\lambda_{k}-\lambda_{m_{3}})^{-1}\nonumber\\
&&\prod_{a_{1}=0}^{N}(a_{1}\lambda_{i}+(N-a_{1})\lambda_{j})\nonumber\\
&&\prod_{a_{2}=0}^{N}(a_{2}\lambda_{i}+(N-a_{2})\lambda_{k})\nonumber\\
&&\prod_{a_{3}=0}^{N}(a_{3}\lambda_{i}+(N-a_{3})\lambda_{l})\nonumber\\
&&\prod_{n\ne i}(\lambda_{i}-\lambda_{n})^{2}\nonumber\\
&&(\mbox{from (g)})
\label{78}
\end{eqnarray}
These results are generically indpendent of the values $\lambda_{i}$,
so we set $\lambda_{i}$ equals to $3^{i}$. Similarly we calculate the
amplitudes $\langle {\cal O}_{e^{\alpha}}{\cal O}_{e^{\beta}}
\rangle_{1}$,$\langle {\cal O}_{e^{\alpha}}{\cal O}_{e^{\beta}}
\rangle_{2}$,$(\alpha + \beta =N-3)$. The results are collected in
{\bf Table 1},$\sim$ {\bf Table 3}.

Note that $\langle {\cal O}_{e^{N-4}}\rangle_{n} =
\langle {\cal O}_{e}{\cal O}_{e^{N-4}}\rangle_{n}\cdot n$.
 This implies that the K\"aler equation of Gromov-Witten
invariants holds
for the amplitudes defined by (\ref{25}). Assuming this relation
for all
amplitudes, the results of  {\bf Table 1}$\sim$ {\bf Table 3} coincides
with the ones calculated from mirror symmetry \cite{gmp}.
\footnote{Note that for
three-point function,amplitudes of the matter theory and the ones of
theory coupled with gravity coincide.}
We can calculate the amplipudes for matter theory for example
$\langle\overbrace{{\cal O}_{e} \cdots {\cal O}_{e} }^{N-2\mbox{times}}
\rangle$ by a rather cunning way. Fusion rules
hold in the matter theory, so we can reduce the amplitudes into
the products of three-point functions.

Consider the ``matter'' expansion
\begin{equation}
\langle {\cal O}_{e}{\cal O}_{e^{\alpha}}{\cal O}_{e^{\beta}}\rangle
= N + \sum_{k=1}^{\infty}
\langle {\cal O}_{e}{\cal O}_{e^{\alpha}}
{\cal O}_{e^{\beta}}\rangle_{k}e^{-kt}
\label{79}
\end{equation}
where $t$ is the deformation parameter coupled to the K\"ahler form.
By using fusion rules, and flat metric $\eta_{\alpha\beta}=
N\cdot \delta_{\alpha+\beta,N-2}$,
\begin{eqnarray}
\langle\overbrace{{\cal O}_{e} \cdots {\cal O}_{e}
}^{N-2\mbox{times}}\rangle&=& N^{5-N}\prod_{i=1}^{N-4}\langle{\cal O}_{e}
{\cal O}_{e^{i}}{\cal O}_{e^{N-3-i}}\rangle\nonumber\\
&=& N + \sum_{k=1}^{\infty}\langle \overbrace{{\cal O}_{e}
\cdots {\cal O}_{e}
}^{N-2\mbox{times}}\rangle_k e^{-kt}
\label{80}
\end{eqnarray}
Then for example,$\langle{\overbrace{{\cal O}_{e} \cdots {\cal O}_{e}
}^{N-2\mbox{times}}}\rangle_{1}$ can be calculated as
\begin{eqnarray}
\langle{\overbrace{{\cal O}_{e} \cdots {\cal O}_{e}
}^{N-2\mbox{times}}}\rangle_{1}&=&
-\frac{1}{2}\sum_{i\ne j}\prod_{k\ne i,j}(\lambda_{i}-\lambda_{k})^{-1}
(\lambda_{j}-\lambda_{k})^{-1}
\prod_{a=0}^{N}(a\lambda_{i}+(N-a)\lambda_{j})\nonumber\\
&&((N-4)(\lambda_{j}^{N-2}-\lambda_{i}^{N-2})-(N-2)
(\lambda_{j}^{N-3}\lambda_{i}-\lambda_{i}^{N-3}\lambda_{j}))\nonumber\\
&&\frac{1}{(\lambda_{i}-\lambda_{j})^3}\\
\label{81}
&=& N^{N+1}-(N-2)\cdot N\cdot N!(\frac{N-1}{1}+\cdots +\frac{1}{N-1})
\nonumber\\
&&-2N\cdot N!
\label{82}
\end{eqnarray}
If we set $\lambda_{i}=i$, we can derive (\ref{81}) from (\ref{82})
by a rather clumsy but elementary calculation. So theoretically
we can see the coincidence of the calculation from A-model and B-model
to the arbitrary degree $n$.

\section{Construction of Free Energy}
In section $4$, we see that we can calculate the amplitudes $\langle
*\rangle_{n,grav.alt.}$ for tpological sigma model on $M_{N}$ coupled to
gravity by torus action method. As we have seen in section $3$ and section $4$
, this method has a structure of summing over tree graphs, so we can
construct a representation of Path-Integral form of the generating
function of all amplitudes, i.e., free energy.

First, let us write out explicitly the contribution from ${\cal M}(\Gamma)$
to the amplitude $\langle {\cal O}_{e^{j_{1}}}\cdots {\cal O}_{e^{j_{k}}}
\rangle_{n,alt.grav.}$
\begin{eqnarray}
\lefteqn{(\mbox{Contribution from ${\cal M}(\Gamma)$ to
$\langle {\cal O}_{e^{j_{1}}}\cdots {\cal O}_{e^{j_{k}}}
\rangle_{n,alt.grav.}$})}\nonumber\\
&&=\frac{1}{{}^{\sharp}Aut(\Gamma)}(\prod_{v\in Vert(\Gamma)}
\lambda_{f_{v}(i)}^{j_{i}}\nonumber\\
&&\prod_{v\in Vert(\Gamma)}\prod_{{flags}\atop {F=(v,\alpha)}}w_{F}^{-1}
(\sum_{{flags}\atop{F=(v,\alpha)}}w_{F}^{-1})^{val(v)
+{}^{\sharp}S_{v}-3}\nonumber\\
&&\prod_{v\in Vert(\Gamma)}(\prod_{j\ne f_{v}}
(\lambda_{f_{v}}-\lambda_{j}))^{val(v)-1}\nonumber\\
&&\prod_{\alpha\in Edge(\Gamma)}\frac{1}{d_{\alpha}}\nonumber\\
&&\prod_{{flags}\atop {F=(v,\alpha)}}(w_{F})^{-d_{\alpha}}
\prod_{{\alpha\in Edge(\Gamma)}\atop{(u,v): vertices\;of\:\alpha}}
\prod_{k\neq f_{u},f_{v}}\prod_{m=0}^{d_{\alpha}}
(\frac{m\lambda_{f_{u}}+
(d_{\alpha}-m)\lambda_{f_{v}}}{d_{\alpha}}-\lambda_{k})^{-1}
\nonumber\\
&&\prod_{\alpha\in Edge(\Gamma)}(d_{\alpha}!)^{-2}
\prod_{v\in Vert(\Gamma)}(\prod_{j:j\neq f_{v}}
(\lambda_{f_{v}}-\lambda_{j}))^{val(v)-1}\nonumber\\
&&\prod_{{\alpha\in Edge(\Gamma)}\atop{(v_{1},v_{2}):
vertices\; of\;\alpha}}
\prod_{a=0}^{Nd_{\alpha}}(\frac{a\lambda_{f_{v_{1}}}+(N-a)
\lambda_{f_{v_{2}}}}{d_{\alpha}})\prod_{v\in Vert(\Gamma)}
(N\lambda_{f_{v}})^{1-val(v)})
\label{83}
\end{eqnarray}
Then we classify the factors into two groups. One is the factors
from edges, and the other from vertices.
The factors from the edges are
\begin{eqnarray}
\lefteqn{\mbox{(i)} \prod_{{flags}\atop{F=(v,\alpha)}}w_{F}^{-1}}
\nonumber\\
&& \mbox{(ii)}\prod_{v\in Vert(\Gamma)}(\prod_{j\ne f_{v}}
(\lambda_{f_{v}}-\lambda_{j}))^{val(v)}\nonumber\\
&& \mbox{(iii)}\prod_{{flags}\atop {F=(v,\alpha)}}(w_{F})^{-d_{\alpha}}
\prod_{{\alpha\in Edge(\Gamma)}\atop{(u,v): vertices\;of\:\alpha}}
\prod_{k\neq f_{u},f_{v}}\prod_{m=0}^{d_{\alpha}}
(\frac{m\lambda_{f_{u}}+
(d_{\alpha}-m)\lambda_{f_{v}}}{d_{\alpha}}-\lambda_{k})^{-1}
\prod_{\alpha\in Edge(\Gamma)}(d_{\alpha}!)^{-2}\nonumber\\
&&\mbox{(iv)}\prod_{{\alpha\in Edge(\Gamma)}\atop{(v_{1},v_{2}):
vertices\; of\; \alpha}}
\prod_{a=0}^{Nd_{\alpha}}(\frac{a\lambda_{f_{v_{1}}}+(N-a)
\lambda_{f_{v_{2}}}}{d_{\alpha}})\prod_{v\in Vert(\Gamma)}
(N\lambda_{f_{v}})^{-val(v)}\nonumber\\
&&\mbox{(v)}\prod_{Edge(\Gamma)}\frac{1}{d_{\alpha}}
\label{84}
\end{eqnarray}
And the factors we can push into the contribution from vertices are,
\begin{eqnarray}
\lefteqn{\mbox{(i)}\prod_{v\in Vert(\Gamma)}
\lambda_{f_{v}(i)}^{j_{i}}}\nonumber\\
&&\mbox{(ii)}\prod_{v\in Vert(\Gamma)}
(\sum_{{flags}\atop{F=(v,\alpha)}}w_{F}^{-1})^{val(v)
+{}^{\sharp}S_{v}-3}\nonumber\\
&&\mbox{(iii)}\prod_{v\in Vert(\Gamma)}(\prod_{j\ne f_{v}}
(\lambda_{f_{v}}-\lambda_{j}))^{-1}\nonumber\\
&&\mbox{(iv)}\prod_{v\in Vert(\Gamma)}(N\lambda_{f_{v}})
\label{85}
\end{eqnarray}
Then we introduce the field variables $\phi_{ij,d}$,propagator
$g_{ij,d;i^{\prime}j^{\prime},d^{\prime}}$, vertex \\
$C_{i_{1}j_{1},d_{1};\cdots
i_{k}j_{k},d_{k}}\phi_{i_{1}j_{1},d_{1}}\cdots\phi_{i_{k}j_{k},d_{k}}$
and external field source parameters $t_{1},\cdots,t_{N-4}$.

In this formulation, field variables correspond to the edges with
characters
i and j and degree d, $g_{ij,d;i^{\prime}j^{\prime},d^{\prime}}$
remains nonzero only if $i=j^{\prime},j=i^{\prime},d=d^{\prime}$, and the
nonzero value of propagator is given as the reciprocal of the product of
(\ref{84}) (i)$\sim$(v). Then we have
\begin{equation}
g_{ij,d}:=g_{ij,d;ji,d}= \frac{-d^{3}(\lambda_{i}-\lambda_{j})^{2}
\prod_{k=1}^{N}\prod_{a=1}^{d-1}(a\lambda_{i}+(d-a)\lambda_{j}-
d\lambda_{k})}{\prod_{a=1}^{Nd-1}(a\lambda_{i}+(Nd-a)\lambda_{j})}
\label{86}
\end{equation}
Vertex  $C_{i_{1}j_{1},d_{1};\cdots
i_{k}j_{k},d_{k}}\phi_{i_{1}j_{1},d_{1}}\cdots\phi_{i_{k}j_{k},d_{k}}$
are constructed with pairing the factor $\lambda_{i}^{k}$ to $t_{k}$
as follows.
\begin{eqnarray}
\lefteqn{\sum_{k=1}^{\infty}\sum_{l=1}^{\infty} \frac{1}{l!}
\sum_{i=1}^{N}
\frac{N\lambda_{i}}{\prod_{j\ne i}(\lambda_{i}-\lambda_{j})}
\sum_{{d_{1},\cdots,d_{k},d_{*}\geq 1}\atop{j_{1},\cdots,j_{k},j_{*}\ne i}}
\sum_{\tilde{j_{1}},\cdots,\tilde{j_{l}} \in \{ 1,2,\cdots,N-4\}}}\nonumber\\
&&(v_{ij_{1},d_{1}}+\cdots+v_{ij_{k},d_{k}})^{k+l-3}
\phi_{ij_{1},d_{1}}\cdots\phi_{ij_{k},d_{k}}
t_{\tilde{j_{1}}}\cdots t_{\tilde{j_{k}}}
(\lambda_{i}^{\tilde{j_{1}}+\cdots +\tilde{j_{k}}})\nonumber\\
&&=\sum_{i=1}^{N}
\frac{N\lambda_{i}}{\prod_{j\ne i}(\lambda_{i}-\lambda_{j})}
\sum_{k=1}^{\infty}\frac{1}{k!}
\sum_{{d_{1},\cdots,d_{k},d_{*}\geq 1}\atop{j_{1},\cdots,j_{k},j_{*}\ne i}}
(v_{ij_{1},d_{1}}+\cdots+v_{ij_{k},d_{k}})^{k-3}
\phi_{ij_{1},d_{1}}\cdots\phi_{ij_{k},d_{k}}\nonumber\\
&&\exp((t_{1}\lambda_{i}+\cdots+t_{N-4}\lambda_{i}^{N-4})
(v_{ij_{1},d_{1}}+\cdots+v_{ij_{k},d_{k}}))\nonumber\\
&&v_{ij,d}:=\frac{d}{\lambda_{i}-\lambda_{j}}
\label{87}
\end{eqnarray}
where $1/k!$ is the factor that produces $1/{}^{\sharp}Aut(\Gamma)$
and $1/l!$ is the combinatorial factor in the insertion of the
external operator.
With these preparation, we have the path-integral representation of
the free energy.
\begin{eqnarray}
\lefteqn{F_{M_{N}}(t_{1},\cdots,t_{N-4})} \nonumber\\
&&:= \sum_{n_{1},\cdots,n_{N-4}\geq 0}\langle {\cal O}^{n_{1}}_{e^{1}}
\cdots {\cal O}^{n_{N-4}}_{e^{N-4}}\rangle\frac{t_{1}^{n_{1}}\cdots
t_{N-4}^{n_{N-4}}}{n_{1}!\cdots n_{N-4}!}\nonumber\\
&&=Res_{z}Res_{{h}}(\frac{1}{z}\log(\mbox{det}(g^{-1})\frac{1}{{h}}
\int d\phi_{ij,d}\nonumber\\
&&(-\frac{1}{2}\sum_{i,j,d}\frac{-d^{3}(z\lambda_{i}-z\lambda_{j})^{2}
\prod_{k=1}^{N}\prod_{a=1}^{d-1}(az\lambda_{i}+(d-a)z\lambda_{j}-
dz\lambda_{k})}{\prod_{a=1}^{Nd-1}(az\lambda_{i}+(Nd-a)z\lambda_{j})}
\phi_{ij,d}\phi_{ji,d}\nonumber\\
&&+\sum_{i=1}^{N}
\frac{Nz\lambda_{i}}{\prod_{j\ne i}(z\lambda_{i}-z\lambda_{j})}
\sum_{k=1}^{\infty}\frac{1}{k!}
\sum_{{d_{1},\cdots,d_{k},d_{*}\geq 1}\atop{j_{1},\cdots,j_{k},
j_{*}\ne i}}
(\frac{v_{ij_{1},d_{1}}}{z}+\cdots+\frac{v_{ij_{k},d_{k}}}{z})^{k-3}
\phi_{ij_{1},d_{1}}\cdots\phi_{ij_{k},d_{k}}\nonumber\\
&&\exp((t_{1}\lambda_{i}z+\cdots+t_{N-4}\lambda_{i}^{N-4}z^{N-4})
(\frac{v_{ij_{1},d_{1}}}{z}+\cdots+\frac{v_{ij_{k},d_{k}}}{z})))))
\label{88}
\end{eqnarray}
where we introduce $h$ and dummy variable $z$ to pick up the
portion that comes from tree graphs and satisfies the topological selection
rule (\ref{18}).
\section{Conclusion}
I believe that results of this paper are reasonably clear, so I just point out
what remains to show, or consider in the future. First, the relation between
the representations of $\langle *\rangle_{n,grav.}$ and
$\langle* \rangle_{n,grav.alt.}$. We think that the situation corresponds to
the
case described by Witten \cite{witten3}, i.e., the zero locus of the
section of
${\cal E}_{Nn+1}$  and external forms are not always points but
submanifolds of ${\cal M}_{0,n,k}^{CP^{N-1}}$. In such cases, Euler classes
on these submanifolds arise and $\langle * \rangle_{n,grav.alt.}$
reduces to $\langle *\rangle_{n,grav.}$. Second, the difference
between the
moduli space of matter theory and the one coupled to gravity.
For the matter theory we asserted in \cite{nj} that moduli spaces
can be constructed
by subtracting boundary points from simple projective space. This
statement
is indirectly supported by the work of Morrison,Plesser \cite{mp}
by use of
gauged linear sigma model. But stable map approach seems to go in
the opposite way, i.e., it adds boundary points to compactify the
moduli space. So the problem of construction of moduli spaces of
matter
theory still remains.\\
For generalization to the worldsheet of higher genera, we can think of two
naive additional approaches. One is the addition of loop amplitudes.
 The
other is the introduction of gravitaional correlation functions for higher
genus in the calculation of (\ref{55}). We tried the calculation of genus
1 amplitudes with these factors but cannot get good results. So,
further
consideration is needed.\\
\smallskip
{\bf Acknowledgement}
\smallskip
I'd like to thank Dr.K.Hori for many useful discussions. I also thank
Dr.Y.Sun,Dr.M.Nagura and Prof.T.Eguchi for kind encouragement.

\newpage
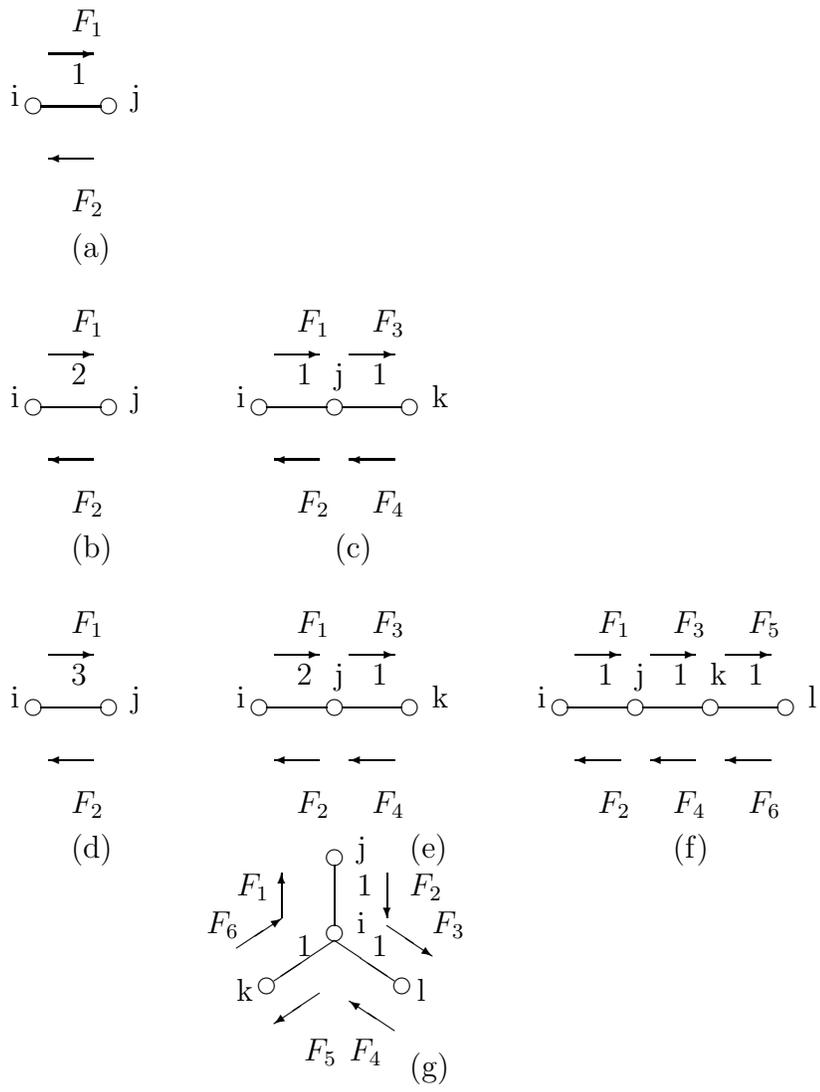
\begin{figure}[p]
\begin{center}
\setlength{\unitlength}{1mm}
\begin{picture}(130,180)
\put(11,130){\line(1,0){8}}
\put(12,137){\vector(1,0){6}}
\put(15,140){\makebox{$F_{1}$}}
\put(15,116){\makebox{$F_{2}$}}
\put(15,110){\makebox{(a)}}
\put(18,123){\vector(-1,0){6}}
\put(7,130){\makebox{i}}
\put(23,130){\makebox{j}}
\put(15,133){\makebox{1}}
\put(11,90){\line(1,0){8}}
\put(15,100){\makebox{$F_{1}$}}
\put(15,76){\makebox{$F_{2}$}}
\put(15,70){\makebox{(b)}}
\put(12,97){\vector(1,0){6}}
\put(18,83){\vector(-1,0){6}}
\put(15,93){\makebox{2}}
\put(7,90){\makebox{i}}
\put(23,90){\makebox{j}}
\put(41,90){\line(1,0){8}}
\put(45,100){\makebox{$F_{1}$}}
\put(45,76){\makebox{$F_{2}$}}
\put(42,97){\vector(1,0){6}}
\put(48,83){\vector(-1,0){6}}
\put(45,93){\makebox{1}}
\put(51,90){\line(1,0){8}}
\put(50,70){\makebox{(c)}}
\put(55,100){\makebox{$F_{3}$}}
\put(55,76){\makebox{$F_{4}$}}
\put(52,97){\vector(1,0){6}}
\put(58,83){\vector(-1,0){6}}
\put(55,93){\makebox{1}}
\put(37,90){\makebox{i}}
\put(50,93){\makebox{j}}
\put(63,90){\makebox{k}}
\put(11,50){\line(1,0){8}}
\put(15,60){\makebox{$F_{1}$}}
\put(15,36){\makebox{$F_{2}$}}
\put(15,30){\makebox{(d)}}
\put(12,57){\vector(1,0){6}}
\put(18,43){\vector(-1,0){6}}
\put(7,50){\makebox{i}}
\put(23,50){\makebox{j}}
\put(15,53){\makebox{3}}
\put(41,50){\line(1,0){8}}
\put(45,60){\makebox{$F_{1}$}}
\put(45,36){\makebox{$F_{2}$}}
\put(42,57){\vector(1,0){6}}
\put(48,43){\vector(-1,0){6}}
\put(51,50){\line(1,0){8}}
\put(60,30){\makebox{(e)}}
\put(55,60){\makebox{$F_{3}$}}
\put(55,36){\makebox{$F_{4}$}}
\put(52,57){\vector(1,0){6}}
\put(58,43){\vector(-1,0){6}}
\put(37,50){\makebox{i}}
\put(50,53){\makebox{j}}
\put(63,50){\makebox{k}}
\put(45,53){\makebox{2}}
\put(55,53){\makebox{1}}
\put(81,50){\line(1,0){8}}
\put(85,60){\makebox{$F_{1}$}}
\put(85,36){\makebox{$F_{2}$}}
\put(82,57){\vector(1,0){6}}
\put(88,43){\vector(-1,0){6}}
\put(91,50){\line(1,0){8}}
\put(95,60){\makebox{$F_{3}$}}
\put(95,36){\makebox{$F_{4}$}}
\put(92,57){\vector(1,0){6}}
\put(98,43){\vector(-1,0){6}}
\put(101,50){\line(1,0){8}}
\put(105,60){\makebox{$F_{5}$}}
\put(105,36){\makebox{$F_{6}$}}
\put(102,57){\vector(1,0){6}}
\put(108,43){\vector(-1,0){6}}
\put(95,30){\makebox{(f)}}
\put(77,50){\makebox{i}}
\put(90,53){\makebox{j}}
\put(100,53){\makebox{k}}
\put(85,53){\makebox{1}}
\put(95,53){\makebox{1}}
\put(105,53){\makebox{1}}
\put(113,50){\makebox{l}}
\put(50,21){\line(0,1){8}}
\put(43,22){\vector(0,1){6}}
\put(57,28){\vector(0,-1){6}}
\put(37,25){\makebox{$F_{1}$}}
\put(60,25){\makebox{$F_{2}$}}
\put(53,30){\makebox{j}}
\put(53,25){\makebox{1}}
\put(50,19){\line(3,-2){8}}
\put(63,20){\makebox{$F_{3}$}}
\put(52,3){\makebox{$F_{4}$}}
\put(46,3){\makebox{$F_{5}$}}
\put(33,20){\makebox{$F_{6}$}}
\put(37,18){\vector(3,2){6}}
\put(48,12){\vector(-3,-2){6}}
\put(58,7){\vector(-3,2){6}}
\put(57,21){\vector(3,-2){6}}
\put(53,20){\makebox{i}}
\put(55,17){\makebox{1}}
\put(37,11){\makebox{k}}
\put(50,19){\line(-3,-2){8}}
\put(61,11){\makebox{l}}
\put(45,17){\makebox{1}}
\put(60,1){\makebox{(g)}}
\put(50,20){\circle{2}}
\put(50,30){\circle{2}}
\put(59,13){\circle{2}}
\put(41,13){\circle{2}}
\put(10,130){\circle{2}}
\put(20,130){\circle{2}}
\put(10,90){\circle{2}}
\put(20,90){\circle{2}}
\put(10,50){\circle{2}}
\put(20,50){\circle{2}}
\put(40,90){\circle{2}}
\put(50,90){\circle{2}}
\put(60,90){\circle{2}}
\put(40,50){\circle{2}}
\put(50,50){\circle{2}}
\put(60,50){\circle{2}}
\put(80,50){\circle{2}}
\put(90,50){\circle{2}}
\put(100,50){\circle{2}}
\put(110,50){\circle{2}}
\end{picture}
\end{center}
\caption{Tree Graphs up to Degree 3}
\end{figure}

\begin{table}[p]
\caption{$\langle {\cal O}_{e^{N-4}}\rangle_{alt,grav}$}
\begin{center}
\begin{tabular}{|l|l|l|l|}
\hline
&$\langle {\cal O}_{e^{N-4}}\rangle_{1}$&
 $\langle {\cal O}_{e^{N-4}}\rangle_{2}$&
 $\langle {\cal O}_{e^{N-4}}\rangle_{3}$\\
\hline
 N=5& 2875& $\frac{4876875}{4}$&$\frac{8564575000}{9}$\\
\hline
 N=6& 60480& 440899200&6255156284160\\
\hline
 N=7& 1009792&122240038536& $\frac{274758045710330728}{9}$\\
\hline
 N=8& 15984640 & 33397163702784 &$\frac{1386812286427888746496}{9}$\\
\hline
 N=9& 253490796 & 9757818404032059 & 897560654227562367535680\\
\hline
 N=10& 4120776000 & 3151991359959750000 & 6298886011657402651840000000\\
\hline
\end{tabular}
\end{center}
\end{table}
\begin{table}[p]
\caption{$\langle {\cal O}_{e^{\alpha}}{\cal O}_{e^{\beta}}
\rangle_{1,alt,grav}$}
\begin{center}
\begin{tabular}{|l|l|}
\hline
 N=5 & $\langle {\cal O}_{e}{\cal O}_{e}\rangle_{1} =
      2875$\\
\hline
 N=6 & $\langle {\cal O}_{e}{\cal O}_{e^{2}}\rangle_{1}=
60480$\\
\hline
 N=7 & $\langle {\cal O}_{e}{\cal O}_{e^{3}} \rangle_{1}=
1009792$\\
& $\langle {\cal O}_{e^{2}}{\cal O}_{e^{2}} \rangle_{1}=
1707797$\\
\hline
 N=8 & $\langle {\cal O}_{e}{\cal O}_{e^{4}}\rangle_{1}=
15984640$\\
 & $\langle {\cal O}_{e^{2}}{\cal O}_{e^{3}}\rangle_{1}=
37502976$\\
\hline
 N=9 & $\langle {\cal O}_{e}{\cal O}_{e^{5}}\rangle_{1}=
253490796$\\
& $\langle {\cal O}_{e^{2}}{\cal O}_{e^{4}}\rangle_{1}=
763954092$\\
& $\langle {\cal O}_{e^{3}}{\cal O}_{e~{3}}\rangle_{1}=
1069047153$\\
\hline
 N=10 & $\langle {\cal O}_{e}{\cal O}_{e^{6}}\rangle_{1}=
4120776000$\\
 & $\langle {\cal O}_{e^{2}}{\cal O}_{e^{5}}\rangle_{1}=
15274952000$\\
 & $\langle {\cal O}_{e^{3}}{\cal O}_{e^{4}}\rangle_{1}=
27768048000$\\
\hline
\end{tabular}
\end{center}
\end{table}
\begin{table}[p]
\caption{$\langle {\cal O}_{e^{\alpha}}{\cal O}_{e^{\beta}}
\rangle_{2,alt,grav}$}
\begin{center}
\begin{tabular}{|l|l|}
\hline
 N=5 & $\langle {\cal O}_{e}{\cal O}_{e}\rangle_{2} =
      \frac{4876875}{2}$\\
\hline
 N=6 & $\langle {\cal O}_{e}{\cal O}_{e^{2}}\rangle_{2}=
881798400$\\
\hline
 N=7 & $\langle {\cal O}_{e}{\cal O}_{e^{3}} \rangle_{2}=
244480077072$\\
& $\langle {\cal O}_{e^{2}}{\cal O}_{e^{2}} \rangle_{2}=
\frac{1021577199083}{2}$\\
\hline
 N=8 & $\langle {\cal O}_{e}{\cal O}_{e^{4}}\rangle_{2}=
66794327405568$\\
 & $\langle {\cal O}_{e^{2}}{\cal O}_{e^{3}}\rangle_{2}=
224340722909184$\\
\hline
 N=9 & $\langle {\cal O}_{e}{\cal O}_{e^{5}}\rangle_{2}=
19515636808064118$\\
& $\langle {\cal O}_{e^{2}}{\cal O}_{e^{4}}\rangle_{2}=
93777295510651590$\\
& $\langle {\cal O}_{e^{3}}{\cal O}_{e~{3}}\rangle_{2}=
\frac{312074853388012521}{2}$\\
\hline
 N=10 & $\langle {\cal O}_{e}{\cal O}_{e^{6}}\rangle_{2}=
6303982719919500000$\\
 & $\langle {\cal O}_{e^{2}}{\cal O}_{e^{5}}\rangle_{2}=
40342298393756700000$\\
 & $\langle {\cal O}_{e^{3}}{\cal O}_{e^{4}}\rangle_{2}=
100290980414189400000$\\
\hline
\end{tabular}
\end{center}
\end{table}

\end{document}